\documentclass[onecolumn,showpacs,amsmath,amssymb,prd,nofootinbib]{revtex4-2}
\usepackage{graphicx}
\usepackage{epsfig}
\usepackage{enumerate}

\def\beq{\begin{equation}}
\def\eeqno#1{\label{#1}\end{equation}}

\def\rarrow{\rightarrow }

\def\dleft{\rlap{{\it D}}\raise 8pt
\hbox{$\scriptscriptstyle\Leftarrow$}}
\def\dright{\rlap{{\it
D}}\raise 8pt\hbox{$\scriptscriptstyle\Rightarrow$}}

\def\msun{M_{\odot}}
\def\az{a_{0}}

\def\l0{\ell_{0}}

\def\rar{\rightarrow}
\def\s{\sigma}

\def\a{\alpha}

\def\c{\gamma}
\def\l{\lambda}

\def\f{\phi}
\def\t{\theta}

\def\k{\kappa}
\def\r{\rho}
\def\m{\mu}
\def\n{\nu}
\def\z{\zeta}

\def\vinf{V\_{\infty}}

\def\A{\mathcal{A}}
\def\B{\mathcal{B}}

\def\o{\omega}

\def\d{\delta}

\def\a{\alpha}
\def\xlimin{{x\rarrow\infty \atop{\raise 1pt\hbox to 30pt
{\rightarrowfill}}}}
\def\limlim#1#2{{#1\rarrow #2 \atop{\raise 1pt\hbox to 30pt
{\rightarrowfill}}}}

\def\vr{{\bf r}}

\def\vR{{\bf R}}
\def\vv{{\bf v}}

\def\vg{{\bf g}}

\def\va{{\bf a}}

\def\vf{{\bf f}}
\def\vF{{\bf F}}
\def\vE{{\bf E}}
\def\vP{{\bf P}}
\def\vJ{{\bf J}}
\def\hro{\hat {\bf  r}(\o)}
\def\hFo{\hat {\bf  F}(\o)}

\def\hao{\hat {\bf  a}(\o)}
\def\hro{\hat {\bf  r}(\o)}

\def\haNo{\hat {\bf  a}_N(\o)}
\def\hFo{\hat {\bf F}(\o)}

\def\ve{{\bf e}}
\def\S{\Sigma}

\def\grad{\vec\nabla}

\def\gf{\grad\phi}

\def\I{\mathcal{I}}

\def\hao{\hat {\bf  a}(\o)}

\def\haNo{\hat {\bf  a}_N(\o)}
\def\I{\mathcal{I}}
\def\Q{\mathcal{Q}}

\def\gN{g\_N}

\def\m{\mu}
\def\a{\alpha}

\def\n{\nu}

\def\_#1{_{\scriptscriptstyle #1}}
\def\^#1{^{\scriptscriptstyle #1}}

\def\RM{r\_M}

\def\oot{\frac{1}{2}}

\def\tpg{2\pi G}
\def\fpg{4\pi G}
\def\epg{8\pi G}

\def\vgN{\vg\_N}

\def\rro#1{\overset{\text{\scriptsize$\leftrightarrow$}}{#1}}

\begin{document}
\title{MOND as manifestation of modified inertia\footnote{Bases on a talk presented at the ``MOND at 40'' conference at St' Andrews University, July 2023.}}

\author{Mordehai Milgrom}
\affiliation{Department of Particle Physics and Astrophysics, Weizmann Institute}

\begin{abstract}
Practically all the full-fledged MOND theories propounded to date are of the modified-gravity (MG) type: they modify only the Newtonian, Poisson action of the gravitational potential, or the general-relativistic, Einstein-Hilbert action, leaving other terms (inertia) intact. Here, I discuss the interpretation of MOND as modified inertia (MI).
My main aim is threefold: (a) to advocate exploring MOND theories beyond MG, and appreciating their idiosyncrasies, (b) to highlight the fact that secondary predictions of such theories can differ materially from those of MG theories, (c) to demonstrate some of this with specific MI models. I discuss some definitions and generalities concerning MI. I then present instances of MI in physics, and the lessons we can learn from them for MOND. I then concentrate on a specific class of nonrelativistic, MOND, MI models, and contrast their predictions with those of the two workhorse, MG theories -- AQUAL and QUMOND. The MI models predict possibly a stronger external-field effect -- e.g. on low acceleration systems in the solar neighborhood -- such as very wide binary stars -- and on vertical motions in disc galaxies. More generally, the workings of the effect are rather different, and depend in different ways on dimensionless characteristics of the system, such as frequency ratios of the external and internal fields, eccentricity of trajectories, etc.
These models predict a {\it much} weaker effect of the Galactic field in the inner Solar System than is predicted by AQUAL/QUMOND. I also show how noncircular motions -- such as those perpendicular to the disc -- modify the standard, algebraic mass-discrepancy-acceleration relation (aka RAR) that is predicted by MI for exactly circular orbits.
These differences, and more that are discussed,  can potentially offer ways to distinguish between theories.
\end{abstract}
\maketitle

\section{Introduction}
MOND \cite{milgrom83} has been propounded as an alternative to dark matter in accounting for the acceleration anomalies in the Universe: The gravitational accelerations calculated in standard dynamics from the observed mass distributions in galactic systems (and the Universe at large) fall very short of the measured accelerations of test particles.
Reviews of MOND can be found in Refs. \cite{fm12,milgrom14,milgrom20,mcgaugh20,merritt20,bz22}.
In particular, Ref. \cite{merritt20} is a detailed account of the MOND research program from the epistemological and philosophical viewpoints.
\par
The MOND research program starts from the following basic tenets: (a) Standard dynamics -- Newtonian and general relativity (GR, hereafter) -- break down in the limit of low accelerations, the limit being defined by a new constant, $\az$, with the dimensions of acceleration. (b) Standard dynamics is restored when all system attributes with the dimensions of acceleration are $\gg\az$ -- or when taking in a MOND theory, or in MOND predictions, the formal limit $\az\rar 0$. (c) Gravitational, nonrelativistic dynamics become space-time scale invariant -- i.e., invariant under $(t,\vr)\rar(\l t,\l\vr)$ -- in the opposite limit, formally approached when taking in a MOND theory, or in MOND predictions, $\az\rar\infty$ (and $G\rar 0$, with $A_0\equiv G\az$ fixed).\footnote{To these tenets one may add the small-print requirement that no fundamental dimensionless constants of values very different from unity appear in the theory. This means, e.g., that $\az$ defines not only where the transition occurs, but also that is occurs within a range of acceleration values of order $\az$. This is a natural requirement and is satisfied, e.g., in the contexts of the relativity and quantum departures from classical Newtonian physics, where $c$ determines everything in the former, and $\hbar$ in the latter.}
\par
From the basic tenets alone follow a large number of predictions; I refer to these as ``primary'', or ``first-tier'' predictions.
Some examples of primary prediction are as follows: (a) The salient properties of disc galaxy rotation curves: asymptotic flatness, the mass-asymptotic-speed relation $V_\infty^4=MG\az$, and, to within small details, the full prediction of rotation curves -- tested in many works; see the above reviews. (b) The ``virial,'' mass-velocity-dispersion relation for deep-MOND (low-acceleration) systems of all types (rotation-, or pressure-supported): $MG\az=\k\s\^4$, with $\k\sim 1$, applied, e.g., in Refs. \cite{mm13,milgrom19}. (c) Mass discrepancies always set in around where the measured acceleration is $\az$ (d) The presence of an external-field effect (hereafter, EFE), whereby an external field, even if uniform, influences the internal dynamics of a gravitating system embedded in it. (e) Definite correlation between the mean baryonic surface density (or gravitational acceleration) of a system (e.g., a galaxy) and that of its putative dark matter halo \cite{milgrom09c,lelli16,milgrom16}. In particular it is predicted that the latter cannot much exceed $\az$ \cite{brada99}. (f) Enhanced stability of galactic discs \cite{milgrom89,brada99a,banik18}. These predictions, and more -- with emphasis on how they follow from only the basic tenets -- are discussed in Ref. \cite{milgrom14a}. References to tests of these predictions can be found in the above mentioned reviews.
\par
As has been stressed repeatedly, there are, however, also phenomena on which the basic tenets themselves do not make absolute and concrete predictions. Different MOND theories -- while embodying the basic tenets, and hence sharing the primary predictions -- may differ on what they predict for these secondary, or lower-tier phenomena.
Among these, one may count the following: (a) Minor details of the rotation curves (e.g., Refs. \cite{brada95,milgrom12,brown18,pl20,chae22}). (b) The exact value of $\k$ in the $M~-~\s$ relation. (c) The exact dependence of the effective two-body force on the masses. (d) The exact workings (and strength) of the EFE. (e)  MOND effects in high-acceleration regions, such as the inner Solar System. (f) The workings of dynamical friction.
\par
Quite a few MOND theories, of varying degrees of development, have been proposed to date. Some are nonrelativistic and underlie MOND phenomenology for self-gravitating nonrelativistic systems. Some are relativistic \cite{bm84,sanders97,soussa03,bekenstein04,zfs07,milgrom09b,deffayet14,sz19,milgrom19a,dambrosio20,sz21,milgrom22}, and reduce to nonrelativistic MOND theories in the nonrelativistic limit, but can vary greatly in their predictions in the relativistic regime.
\par
Notable among the nonrelativistic theories are the AQUAL (aquadratic Lagrangian) \cite{bm84}, and the quasilinear MOND theory (QUMOND) \cite{milgrom10}. The recently proposed tripotential theories (TRIMOND) \cite{milgrom23b} comprise a larger class that includes AQUAL and QUMOND as special cases.
And, QUMOND itself has been generalized in ways that go beyond the basic tenets of MOND (in that they involve dimensioned constants other than $\az$) \cite{milgrom23a}. All these are MG theories.
\par
AQUAL and QUMOND have become the main workhorses, and have been used in deriving various MOND predictions, and for solving more involved problems numerically. Just a few of the many applications of these two theories are described in Refs. \cite{bm84,milgrom86,brada95,brada00,brada00a,tiret08,milgrom10,candlish15,thomas17,bilek17,banik22,banik22a,chae22}).
The two theories tend to make very similar predictions, differing only in minor details.
\par
It seems that as the result of their ubiquitous and exclusive use, the notion has taken root that these twin theories encompass the full scope of MOND, and that all their predictions are universal MOND predictions.
\par
{\it This, however, if far from true, and one of the aims of this contribution is to bring this fact home forcefully.}
\par
The departures of AQUAL and QUMOND from Newtonian gravity are essentially encapsulated in a single ``interpolating function'' that is introduced by hand in their Lagrangian, and which thus dictates all the predictions of these theories.
But this need not be the case even within the class of MG. Other MG theories can make different predictions on secondary phenomena. This was discussed and demonstrated, e.g., in Refs. \cite{milgrom23b,milgrom23a}. For example, these theories involve interpolating functions of several variables, and show a more complicated interpolation scheme between the Newtonian and the deep-MOND regime.
\par
But an even more pronounced variety in second-tier predictions appears when we consider theories that go beyond MG -- I call these modified-inertia theories, because they depart from the Newtonian law of inertia (second law). Such was, in fact, my initial view of MOND.
\par
There are clear and immediately recognized differences between MG and MI. One clear difference is that MI may entail departures from standard dynamics also when forces other than gravity affect the motions (see e.g., Ref. \cite{milgrom15a}). However, since all the information on the acceleration anomalies has so far come from purely gravitating systems, we have not yet been informed observationally on this possible aspect of MOND. But, such differences can in principle be demonstrated with the following two (gedanken?) examples. In the first, only MI predicts an anomaly, and in the other only MG does.
\par
An experiment has been proposed in Ref. \cite{ignatiev07}: For two very brief time intervals every year, in very small regions attached to the Earth's surface, the acceleration due to the revolution of Earth around the Sun cancels that of the rotation of the Earth. The latter acceleration involves also nongravitational forces that keep the experiment attached to the surface.
While the external kinematic acceleration can be very small for these small regions of space-time, the gravitational acceleration is still very large. Thus, an anomaly is expected only in MI.
\par
Another experiment has been proposed, which employs the regions around the saddle points between the Sun and a planet, where the net gravitational acceleration of all bodies (including the Galaxy) is $\lesssim\az$ (see, e.g., Refs. \cite{magueijo06,magueijo12,penner20,bz22}). One strategy could be to have a spacecraft fly by such a region with sensitive accelerometers on board, and make measurements. For this small region of space-time, the spacecraft is freely falling in a sub-$\az$ gravitational field, and an anomaly is expected in any MOND theory.
But in another procedure, the spacecraft is made to follow the saddle point, so as to spend more time in the low-gravitational-acceleration region. This has to be done by nongravitational forces. While an anomaly is expected in MG theories, practically none is expected in MI, because then the kinematic acceleration of the spacecraft is much larger than $\az$.
\par
Another immediate and phenomenologically-important difference between MG and MI is the following: In studying the dynamics of galactic systems, the main tool is measuring kinematic accelerations ($\va=\ddot\vr$) of test particles in the system. In MG, Newton's 2nd law, which remains intact in MG, tells us directly the gravitational acceleration ($\vg=\vF/m$) on the particles, which we can proceed to analyze according to various MG theories. In MI, the directly-measured acceleration does not equal $\vg$, and does not tell us what the force is. In fact, the accelerations $\va$ of different particles {\it at the same position} in a gravitational force field may be different, depending on the particle's trajectory.
\par
And, there are many other, perhaps more subtle, differences between the predictions of MG and MI theories.
Such differences are one of the main topics of this paper.
\par
For many years, we had been testing only the primary predictions of MOND, and it had been less pressing perhaps to recognize such potential differences in secondary predictions.
But in recent years we have begun to be able to test secondary predictions. For example: details of the EFE and how it affects the outskirts of disc galaxies \cite{wu15,hees16,haghi16,chae20,chae21,stiskalek23,desmond23}, the dynamics in wide-binary-stars \cite{hernandez12,bps21,chae23,hernandez23,chae23a,hernandez23a,banik23}, and the appearance of tidal tails of star clusters \cite{thomas18,kroupa22}. Another example of a test of secondary predictions that is becoming feasible is looking for the possible small effect of the galactic acceleration on the motions of bodies in the inner Solar System \cite{milgrom09a,novak11,hees16}. Yet another such second-tier issue (related somewhat to the details of the EFE) is the dynamics perpendicular to the disc in spiral galaxies \cite{bienayme14,loebman14,angus15,milgrom15,angus16,milgrom18,lisanti19}.
\par
Clearly, it is thus important that we understand which theories are constrained by which experiment or observation, and, in particular, to assess possible differences between MI and MG.
But, besides my own preliminary proposals on the subject \cite{milgrom94,milgrom99,milgrom02,milgrom06,milgrom11,milgrom22a}, there have been, as far as I am aware, only very few attempts -- not very promising in my judgement -- to formulate MOND as MI \cite{wang10,namouni15,alzain17,cfp19}.
\par
There is also a more general motivation for casting one's net wider than the existing MOND theories. Several previous papers (e.g., the above-mentioned reviews) discuss the shortcomings of existing MOND theories, the fact that -- very useful as they are -- they can only be approximate, effective theories, and the need to find a ``FUNDAMOND'' a more fundamental theory for MOND. (This is also the state of affairs with other, more entrenched physical theories -- such as the standard model of particle physics.)
\par
Studying and understanding a wider range of effective theories, such as the various MI effective theories, may eventually point the way to such a more fundamental basis for MOND, or to the understanding of how MOND phenomenology emerges from a microscopic picture.
\par
In Sec. \ref{whatismi}, I describe what is meant by MI, and discuss some of the reasons why, I think, the development of MI, MOND theories has not kept pace with that of MG theories. Section \ref{examples} brings some familiar and less familiar examples of modified, or acquired, inertia, to demonstrate and urge that this direction may be quite promising. In Sec. \ref{mondcontext}, I raise some generalities regarding MI in the specific context of MOND. Section \ref{lowertier} lists and details some MOND predictions that are expected to differ among MOND theories -- so-called lower-tier predictions -- because they do not follow robustly from the basic tenets of MOND.
Section \ref{models} describes a class of MI models, which I use to demonstrate some of the possible variety in lower-tier predictions.
Section \ref{discussion} is a discussion.

\section{What is modified inertia?  \label{whatismi}}
It is usually straightforward to define MG in the context of modifying Newtonian dynamics or GR. It would be a modification that disappears in the absence of gravity, or in describing systems where gravity (which is always present) can be neglected compared with other influences; this limit can be formally effected by takeing $G\rar 0$ in the theory. In such a limit, MG leaves us with standard Newtonian dynamics and special relativity, respectively. More specifically, in Newtonian dynamics, MG is a theory that modifies the Poisson term in the Lagrangian (see below). And, MG of GR is a theory that can be formulated as a metric theory that leaves intact all the couplings of matter degrees of freedom (DoFs, hereafter) to the metric (so called ``minimal coupling''), but modifies the Einstein-Hilbert action.\footnote{There are theories, such as TeVeS \cite{bekenstein04}, that can be formulated in equivalent ways; one as just described, and the other in terms of a metric whose action is the Einstein-Hilbert one, but its coupling to matter is modified. These are still MG theories.}
\par
But how to define MI?
In all the  MI, MOND theories proposed to date -- which are all nonrelativistic, and concern only particle dynamics (as opposed to dynamics of fields) --  the concept is well defined (see below), and involves modifying the kinetic particle Lagrangian $mV^2/2$, or the ``inertia term'' $m\va$ in the particle's equations of motion.
But how to define MI more generally?
One can simply view it as a generic name for any theory that is not restricted to modifying gravity in the above sense.
But, perhaps more proper for the name, one can define MI as a theory that modifies the ``free'' or ``kinetic'' terms in the action underlying the theory, or in the equations of motion (or field equations). These are the terms that in the absence of gravity depend on only one DoF (possibly tensorial). Terms that depend on more than one degree of freedom are ``interaction terms''. And gravity itself couples to all matter degrees of freedom.
\par
The statement that the free terms in the action encapsulate inertia can be justified, for example, by noting that the isolated contribution of a DoF to the energy-momentum tensor can be derived from its free action. So, these terms describe the energy-momentum cost of the different configurations of this DoF, and tell us how much energy-momentum have to be invested or extracted to change configurations of the DoF -- the essence of what inertia is.\footnote{In particle dynamics, the free term tells us what the energy and momentum are of a particle with a given speed.}
\par
In a more ``physical'' than formal picture, I envisage inertia -- hence the free actions -- as resulting from what, in a more fundamental theory, are interactions of the DoFs with some inertia-endowing medium, call it the ``inerton''.
In an extreme, ideal picture, one would have a theory with only interactions between the ``known'' DoFs, which also interact with the inerton
(in Ref. \cite{milgrom99} I suggested that the ``inerton'' is the quantum vacuum). Elimination of the inerton DoF from the theory reduces these latter interaction to approximate ``free actions'' for the different DoFs.
\par
In the restricted treatment of MI as a modification of the particle inertia, we leave intact the forces to which the particle is subject, including gravity.
However, it is quite conceivable that in any picture of MI, all free actions including those of all fields, are modified.
For example, if MI is to underlie MOND, the acceleration anomalies exhibited by gravitational lensing by galactic systems, tell us that the inertia of photons, hence the (Maxwell) free action of the electromagnetic field, must also be modified. Also, the Einstein-Hilbert, GR action, or the Poisson action in Newtonian dynamics, may be viewed as the ``free actions'' of gravity. So it is quite conceivable that they too are modified in an ultimate theory of MOND MI.
\par
Thus, to justify our restricted treatment of MI, we need to assume that there is a level of approximation which applies to our case of interest, and in which the force field -- inasmuch as they are static -- are not modified; only the action of the particle, whose DoFs are time dependent needs to be modified. In other circumstances, such as when considering the propagation of electromagnetic waves, or of gravitational waves, the appropriate modification needs to be applied to their inertia as well.
\par
Such an approximation of selective, or stepwise modification, is common in other instances of modified physics; e.g., when we treat the quantum mechanics of quantum particles in static external fields. We do so, for example, in describing atoms by the (nonrelativistic or relativistic) quantum mechanics of electrons in the static and non-quantum Coulomb field of a nucleus. But ultimately we also need to treat the fields themselves quantum mechanically. For example, we need to treat the electromagnetic field as quantized in the context of black-body radiation, or the photoelectric effect.\footnote{In the case of atoms such an extension was forced, for example, by the discovery of the Lamb shift, unaccounted for by the Dirac equation, which treats the electromagnetic field that the electrons see as non-quantum.}
\par
Another example of such a stepwise modification is the stepping stone to quantum gravity, where one describes quantum-gravitational phenomena by quantum field theory on an unquantized background curved space time.
\par
More concretely now,
in this review, and in my previous papers on MI in the context of nonrelativistic MOND \cite{milgrom94,milgrom99,milgrom02,milgrom06,milgrom11,milgrom22a}, I concentrated on the most restricted meaning of MI, in which only the nonrelativistic, kinetic, particle action is modified into a more general functional of the particles trajectory $\vr(t)$:
Start then with the Newtonian action governing the dynamics of a nonrelativistic, self-gravitating, system of (point) masses, $m\_p$, on trajectories $\vr\_p(t)$, which is $I=\int L~dt$, where the Lagrangian, $L$, is
\beq L=-\frac{1}{\epg}\int d^3x~(\gf)^2 +\sum_i m\_p[\oot \vv\_p^2-\f(\vr\_p)],   \eeqno{newtlag}
The equations of motion of the masses and the field equation for the gravitational potential are
\beq \ddot\vr\_p=-\gf(\vr\_p),~~~~~~~~~~~~\Delta\f=\fpg\rho,~~~~~{\rm where}~~~~~~~\r(\vr,t)=\sum_i m\_p\d[\vr-\vr\_p(t)].   \eeqno{equat}
The MG, MOND theories that have been propounded to date (AQUAL \cite{bm84}, QUMOND \cite{milgrom10}, TRIMOND \cite{milgrom23b}, GQUMOND \cite{milgrom23a}, and the more general theories treated in Ref. \cite{milgrom14b}) modify the first term in expression (\ref{newtlag}). Then, only the left-hand side of the Poisson equation is modified to provide a modified prescription for calculating the gravitational potential.
Particles are accelerated by this potential in the standard way.
\par
Modified inertia in the restricted sense I adopt here modifies the particle kinetic term in the action, $m\_p\int\oot \vv\_p^2~dt$ (or the corresponding term in the equation of motion), replacing it with some functional of the particle trajectory of the form $m\_p F[\vr\_p(t),\az]$ (or a generalized inertia functional of the trajectory $m\_p{\bf I}[\vr\_p(t),\az]$). The proportionality to $m\_p$ is retained so as not to violate the universality of free fall in a gravitational field. In a MOND theory, the action functional depends on the MOND constant, $\az$.
\par
Note that this treatment already introduces some ambiguity in the definition of MI. The above Newtonian action is the nonrelativistic limit of the  standard, relativistic, free kinetic action for a particle of mass $m$, which is $-mc^2\int d\tau$. But the nonrelativistic limit of this action is $m\int(\vv^2/2-\f)dt$, where $\f$ is the nonrelativistic gravitational potential. Namely, in the presence of gravity, the relativistic free action gives both the Newtonian particle free action and its interaction with the potential.
\par
It is yet to be seen whether our modification of only the nonrelativistic particle action can result as a nonrelativistic approximation from some more fundamental theory of MI.

\subsection{Who is afraid of modified inertia?  \label{afraid}}
The interpretation of MOND as MI was on the table from the very inception of MOND \cite{milgrom83}. Yet, very little work has been done in this promising vein since then -- far less than on developing MG theories, and on studying their implications analytically and numerically.
\par
Why is it so?
\par
It is true that MI theories seem to entail more drastic modifications of standard dynamics, and there is a natural tendency to minimize change. To boot, there has been much work done already on modifying general-relativistic gravity from the very advent of GR itself. So, from what I have heard, developers feel much more comfortable with searching and propounding MG theories.
It is also true that it has proven rather hard to incorporate the idea of MI in full-fledged theories, based on the standard requirement -- symmetries, conservation laws, etc. All this seems to have kept developers away.
\par
And, it is also true that the MI theories that have been put on the table (including the ones in Ref. \cite{milgrom22}, and which I will discuss below) are rather less amenable to application. Far fewer predictions have been extracted from such formulations, and they are hard  to simulate  numerically. This seems to have kept potential users away.
\par
But all these should not argue against the MI approach, which is otherwise very promising.
General relativity is much harder to solve and apply than Newtonian gravity, and quantum mechanics, to say nothing of quantum field theory, is much harder to solve and apply than classical mechanics. The processes of development, understanding, interpretation, and devising approximation methods to solve and apply these more advanced theories, has been very tortuous, and is still continuing.
\par
Partly to allay misgivings about MI, I now point out that some well established theories may be viewed as MI with respect to Newtonian inertia, even though they are not usually described or thought of as such. I also pinpoint some lessons we can learn from these about what we can expect from MI in the context of MOND.

\section{Familiar examples of modified inertia  \label{examples}}
\subsection{Special relativity as modified inertia}
The departure of special relativity from classical dynamics, entails a modification of Newtonian inertia.
Its kinetic (free) action for a particle of rest mass $m$ is $-mc^2\int d\tau=-mc^2\int [1-(\vv/c)\^2]^{1/2}dt$ ($\tau$ is the proper time), which modifies the Newtonian kinetic action $m\int(\vv\^2/2)~dt$. This gives an inertia force of the form $md(\c\vv)/dt$, instead of the Newtonian $md\vv/dt$ ($\c$ is the Lorentz factor).
The particle equation of motion is
\beq  \frac{dp\^\a}{d\tau}\equiv m\frac{d\^2x\^\a}{d\tau\^2}=f\^\a,  \eeqno{reteqat}
where $f\^\a$ is the relativistic force four-vector. Write the space part of this equation as
\beq  \frac{dp\^i}{dt}=\c\^{-1}f\^i\equiv F^i,  \eeqno{retrty}
where $p\^i=m\c dx\^i/dt$.  For example, for a particle of charge $e$ in an electric field $\vE$, we have $\vF=e\vE$.
\par
We can write this as
\beq \frac{md(\c\vv)}{dt}\equiv m\rro{\m}\left(\frac{\vv}{c}\right)\va=\vF,  \eeqno{mayata}
which can be inverted to read
\beq   m\va=\rro{\n}\left(\frac{\vv}{c}\right)\vF.  \eeqno{pushter}
I wrote these expressions using the ``interpolation matrices'' $\rro{\m}$ and $\rro{\n}$,  defined as
\beq \rro{\m}\left(\frac{\vv}{c}\right)\equiv\c (1+\c^2\frac{\vv\otimes\vv}{c^2}), ~~~~~~\rro{\n}\left(\frac{\vv}{c}\right)\equiv \rro{\m}\^{-1}\left(\frac{\vv}{c}\right)= \c^{-1}\left(1-\frac{\vv\otimes\vv}{c^2}\right), \eeqno{damure}
to resemble MOND relations -- with $\m(x)$ appearing in AQUAL, and $\n(y)$ in QUMOND.
In the Newtonian limit, $c\rar\infty$, $\rro{\n}$ and $\rro{\m}$ tend to the unit matrix.
\par
The Lorentz factor itself is another  relativistic ``interpolating function'' between the Newtonian $|\vv|\ll c$ regime, and the relativistic $|\vv|\approx c$ regime. It appears, e.g., in the expression of time dilations.

\subsubsection{Some lessons}
There are some important lessons that we can learn when comparing (special and general) relativity with Newtonian-dynamics.
\par
In astronomy -- e.g., when probing galaxy dynamics governed by gravity -- we measure kinematic quantities such as velocities and accelerations.
In Newtonian dynamics, the measured kinematic accelerations, $\ddot\vr$ are equal to, and thus give us directly, the gravitational forces (per unit mass) which we are after. This is also the case in nonrelativistic, MG MOND theories such as AQUAL, QUMOND, TRIMOND, and GQUMOND.
\par
But, {\it this need not be the case in other theories} -- as indeed is not the case in MI MOND theories. In general, forces per unit mass are not equal accelerations. In momentum-conserving theories, they are equal the rate of change of some momentum, which we do not measure directly, and whose relation to the measured kinematic quantities, such as acceleration, depends on the specific theory.
\par
This, as we saw, is well exemplified by special relativity.
\par
As a result, intuitive conclusions that are drawn based on our familiarity with Newtonian dynamics can be quite misleading when other theories are at work:
\par
A particle can have very different kinematic accelerations at the same point in a given force field, depending on other of its orbital characteristics.
In special relativity, for example, for
circular orbits:  $\rro{\m}\va=\c\va$, while for linear trajectories $\rro{\m}\va=\c^{3}\va $. More generally, the acceleration of an electron  in an electric field depends not only on their position, but also on their velocity vector.
\par
In particular, the kinematic acceleration, $\ddot\vr$, is not necessarily parallel to the force. And,
no {\bf acceleration} field is defined by a force field, only a $\frac{d\vP}{dt}$ field.
\par
Thus, if such a theory underlays the dynamics, trying to map the force field with different test particles assuming the $\va\propto \vF$, can give different results.
\par
The kinetic, special-relativistic action does not have Galilean invariance. It does have, however, the observationally-all-important boost invariance in the form of Lorentz invariance, generalizing Galilei invariance. It may be that MOND inertia is also underlaid by a new boost symmetry generalizing Lorentz invariance. More on this possibility in Ref. \cite{milgrom06}.
\par
As regards interpolating functions, as they appear in relativity, some lessons are:
(a) The interpolating functions are not introduced ``by hand'' (in special relativity they are dictated by Lorentz invariance).
(b) Different interpolating functions can underlie different phenomena: Besides $\rro{\m}$, and the Lorentz factor discussed above, which are functions of $\vv/c$, in GR (which is modified Newtonian gravity), we have other interpolating functions describing, e.g., the transition from strong gravity near a relativistic object, to the asymptotic, Newtonian gravity, which can be thought of as a function of $MG/Rc^2$.
(c) Interpolating functions, when they appear in relativity, have, as independent variables, quantities of the form $q/c$, where $q$ is a system attribute with the dimensions of velocity, and there may be many such variables, such as $\vv/c$ and $MG/Rc^2$ in the examples above.
\par
All this is to be contrasted with the situation in theories such as AQUAL and QUMOND, where there is a single interpolating function, of a single variable, $|\gf|/\az$, introduced by hand at the level of the action, and thus appearing in all the predictions of this theories.
In TRIMOND and GQUMOND the interpolating functions can depend on several variables. This is also the case for the MI models described in Sec. \ref{models} below.
\par
Finally, we learn from the history of the theory of relativity that, as expected, one ultimately cannot stop at modifying only particle inertia, and leave the force fields themselves untouched, although this can be a very good approximation in some instances.

\subsection{Quantum mechanics as modified inertia}
Consider the quantum mechanics of particles in a given, unquantized force field, such as of electrons interacting electrostatically among themselves and with an external field. This approximation --  referred to sometimes as ``First quantization'' -- works very well for the description of electrons in atoms, or in solids, for example.
\par
In quantum mechanics, the departure of particle dynamics from the Newtonian description is even further reaching than in relativity:
It is not a mere modification of the law dictating how the particle position $\vr(t)$ varies with time. The very description of the particle's whereabouts and motion is modified.
For a many-body system, the Newtonian description of the system's history in terms of $\vr_1(t),\vr_2(t),...$ is replaced by the system's
wave function
$\psi(t, \vr_1,\vr_2,...)$. The positions and velocities of the particles cannot be specified at the same time. Only probabilities are determined by the theory. Particles have internal degrees of freedom  -- their spin and its components, and spin-statistics limits the domain of allowed many-particle wave function. Particles can tunnel through classically impenetrable barriers. The evolution of the wave function is dictated by the Schr\"{o}dinger equation. All these have no parallels in Newtonian dynamics.

\subsubsection{Some lessons}
Some of the message we may take from quantum mechanics as modified Newtonian dynamics are as follows:
MOND too could be much more drastic a modification than just adding gravitational-field degrees of freedom, or changing the field equation of the gravitational fields, or modifying the particle equations of motion.
\par
Regarding interpolating functions.
As in MOND, the quantum mechanical description of various phenomena is encapsulated in ``interpolating functions''. These are -- as in the case of MOND -- quantum mechanical expressions that are functions of variables of the form $q/\hbar$, where $q$ has the dimensions of $\hbar$; so it could be an angular momentum, an action variable, etc. These expressions interpolate between the Newtonian regime -- achieved in the formal limit $\hbar\rar 0$ -- and the quantum regime.
\par
An example is the black-body function, of the variable $q=kT/\o$, which interpolates between the classical Rayleigh–Jeans law, for $q\gg\hbar$, and the Wien law for $q\ll\hbar$. Another example is the quantum mechanical expression for the specific heat of solids -- modelled first by Einstein and then improved on by Debye -- interpolating between the classical Dulong-Petit expression at high temperature and the quantum behavior at low temperatures.\footnote{Einstein's model invoked some material-characteristic frequency, $\o\_0$, and the crucial variable became $q=kT/\o\_0$. Debye's model involved a continuum of characteristic frequencies.} Yet another examples is the expression for the barrier penetration probability, where the crucial variable is $q=d[m(V_0-E)]^{1/2}$ (d is the characteristic width of the barrier and $V_0$ its height, while $m$ is the mass of the tunneling particle, and $E$ its energy.)
\par
In contradistinction with the interpolating functions that appear in theories like AQUAL and QUMOND, those appearing in quantum mechanics are not introduced by hand at the fundamental level, they are derived expressions; are not of a single, universal form that applies to all the phenomena; and, are not necessarily functions of a single variable, or even of the same variable. A single variable appears in the above examples, but it may well happen that more than one crucial variable appears, for example, in systems with more than one characteristic frequency -- as, e.g. in the description of molecules characterized by rotational-vibrational frequencies describing roughly the motion of the nuclei, on one hand, and the electronic frequencies on the other.
\par
Theories such as GQUMOND and TRIMOND were designed to demonstrate that even in MG it is possible to have more than one variable appearing. We shall see that this is also the case in the MI models discussed below.
It may even happen that some of the variables are in one limit (e.g. Newtonian, $q/\az\gg 1$) and others are in the deep-MOND limit ($q/\az\ll 1$), just as this can happen in the quantum case.
\par
Another lesson comes from our knowledge that we cannot stop at quantum mechanics (``first quantization''). The force fields must also be consistently quantized, leading to quantum field theory in flat space-time, extended to the approximate description of quantum fields in a classical curved space-time, and to the yet-to-be-achieved quantization of space-time itself.
\par
Lastly, a lesson regarding the degree to which theories lend themselves to the penetration, manipulation, and solution by us, humans.
MOND theories are rather less wieldy than the simple and linear Newtonian dynamics. They are more difficult to solve, and in some cases it is even hard to see how they can be solved for general N-body systems. Contrary to what I have heard occasionally, our difficulties, or technical inability to solve such problems do not, in any way, argue against the validity of such theories in describing nature.
We know that the similar  difficulties arise even in quantum mechanics, the first approximation in the quantum development, to say nothing of the higher stages, such as quantum-field theory (e.g., the standard model of particle physics). And the same is true of the theory of GR. It has taken many years to develop even crude approximation methods to many of the problems that arise in such theories.
\subsection{Other instances of modified or acquired inertia in physics}
The idea that inertia is an acquired property is old, generally going under the name of ``Mach's principle''. According to it, inertia is due to the interaction of bodies with some omnipresent medium. Ideas along this line differ on what the medium is, and on the nature of the interaction.
But if indeed such is the origin of inertia, then inertia is not some property of ultimate fundamentality, but can take different forms depending on where we are in parameter space of the body with respect to signposts and boundary stones defined by the characteristics of the medium.
\par
For example, I suggested in Ref. \cite{milgrom99} that the quantum vacuum is the omnipresent medium, the interaction with which endows bodies with inertia. This was shown to be appealing in the context of MOND, because the quantum vacuum is imprinted with some characteristics of the global state of the Universe. For example, in a de Sitter cosmological background, the de Sitter radius, $\ell\_\Lambda$, enters the Gibbons-Hawking temperature, and the Unruh temperature of accelerated systems. This idea resonates well with MOND because of the numerical relation of this radius to $\az\sim c^2/\ell\_\Lambda$ \cite{milgrom99,milgrom20a}.
\par
Examples of acquired or modified inertia are very common in physics.
In the standard model of particle physics, the masses of the leptons and quarks (``fundamentally massless'') are acquired by their interaction with the omnipresent Higgs field, each according to the strength of its interaction. The effective inertia of the
electroweak mediators as they are measured in the laboratory also appear in ``low-energy'' circumstances. The inertia of all these would look very different under other circumstances (with the same physics).
\par
Mass renormalization in quantum-field theory is another example, where the interaction with the vacuum changes dramatically the effective masses of particles.
\par
Electrons in solids attain a dispersion relation -- a relation between energy and momentum -- that can differ greatly from the relation $E\_K=P^2/2m$, which they obey when free, $m$ being the electron (inertial) mass.
In some solids and in some regions of phase space they still behave like free electrons but obey this relation with a very different mass.
\par
\begin{figure}[!ht]
\includegraphics[width=.5\columnwidth]{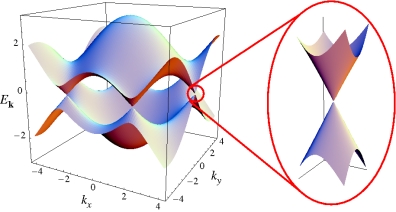}
\caption{Dirac cones in graphene: The kinetic-energy-vs-momentum surfaces of electrons in graphene. It is characterized by cones that are analogue to light cones -- only with an effective ``light speed''$\ll c$, around which electrons behave like massless particles. }\label{dirac}
\end{figure}

It has been observed that in some regions of phase space, electrons in certain systems/materials, such as graphene, behave as massless, dirac particle, with a dispersion relation $E\_K\propto |\vP|$ (the magnified region in Fig. \ref{dirac}). We can also see in Fig. \ref{dirac} that the general ``inertial law'', $E\_K(\vP)$, is more complicated, and takes a different form in other regions of phase space (${\bf k}$ in the figure is the momentum in units of $\hbar$).
\par
Ab initio calculations of the inertia of the electron (due essentially to its electrostatic energy) were popular around the turn of the 20th century, before special relativity took a firm hold in the mind. There were, e.g., the results of Abraham, of Bucherer, and of Lorentz, which Kaufmann set to test experimentally \cite{cushing81}.
\par
In Ref. \cite{milgrom06a}, I showed how certain objects, weakly interacting with a compressible, irrotational, nonviscous fluid, acquire effective inertia, as well as interacting ``gravitationally'' among themselves. They have no free Lagrangian to star with, but the interaction with the fluid endows them with a ``free'' Lagrangian of the form $Mc^2\ell(v^2/c^2)$, where $M$, the acquired ``rest mass'' is a constant that depends on the structure of the body, $c$ is the fluid's speed of sound, playing the role of the speed of light, and $v$ is the velocity of the body with respect to the fluid. As in relativity, $\ell$ diverges as its argument approaches 1. Interestingly, in three dimensions this acoustic analog of inertia gives the same inertia as derived by Abraham and tested by Kaufmann a hundred years earlier, and in a different setting.
In four dimensions, the Lagrangian has a form nearly that of special relativity.
\par
There are many other instances of acquired inertia within well accepted physical theories.

\section{Restricted modified inertia in the context of MOND \label{mondcontext}}
Let us then confine our subsequent discussion to nonrelativistic MOND theories that can be described at MI in the restricted sense defined in Sec. \ref{whatismi}. Such theories are required to retain the standard conservation laws of momentum, energy, and angular momentum. One should also retain some kind boost invariance. Without it, there would be a meaning to absolute overall velocity of a system, which will affect the system's internal dynamics: two similar systems could have different internal dynamics if they move with respect to each other, which we wish to avoid.
\par
These requirements are automatically ensured if the modification involves modifying the Newtonian particle action $I\_K=\int \frac{1}{2}m\vv^2 ~dt$ in a way that the standard symmetries (space and time translations, rotation, and Galilei) are retained.
One can also modify directly the Newtonian particle-inertia term in the equation of motion, $m\va$, in a way that retains the above requirements.
\par
In Ref. \cite{milgrom94}, I showed that a nonrelativistic, MOND, kinetic action for a particle, with the above requirements, and the boost symmetry being Galilean, has to be time nonlocal. Namely, it cannot be of the form $\int L~dt$, where the Lagrangian $L$ is a function of a finite number of time derivatives of the particle trajectory $\vr(t)$.
\par
This follows from the result, derived in Ref. \cite{milgrom94}, that a Galilei-invariant, MOND-based, local action, with a high-acceleration Newtonian limit, has to have a Lagrangian of the form
\beq  L=\oot\vv\^2+L^*(\az,d^2\vr/dt^2,...,d^n\vr/dt^n),   \eeqno{lalaga}
which cannot have a correct MOND limit for $\az\rar 0$.
\par
An example of a simplistic attempt that fails the above criteria was the pristine MOND equation of motion \cite{milgrom83} of a particle subject to a Newtonian gravitational acceleration, $\vgN$, which replaced the Newtonian equation, $\va=\vgN$, by  $\m(a/\az)\va=\vgN$. The inertia term $\m(a/\az)\va$ is translation and rotation invariant, and is also Galilei invariant. However, it leads to nonconservation of momentum in a many-body system. The symmetries do not ensure conservation laws if the equation of motion is not derivable from an action -- as is the case here.
This is why in Ref. \cite{milgrom83}, I had to limit its applicability to test particles.
\par
In another simplistic, but rejectable, example, the kinetic action
\beq  I\_K=\int \frac{1}{2}m\Q(|\dot \vv|/\az)\vv^2 ~dt,  \eeqno{kinact}
is local, obeys the standard symmetries and hence defines conserved momentum, energy, and angular momentum. Its (fourth-order) equation of motion also predicts the standard, algebraic MOND relation, $\m(a/\az)a=g\_N$, for rotation curves {\it as measured in a frame where the velocity of the center of a galaxy vanishes}. But this action has no boost invariance.
\par
As mentioned above, perhaps we can find a local, MI, MOND theory that does enjoy a boost symmetry other than Galilei -- just as special relativity replaces Galilei with Lorentz boost invariance. This possibility was discussed in more detail in Ref. \cite{milgrom06}.
Such a theory has, however, proven difficult to find for me. For this reason, all the MI models I discussed in Refs. \cite{milgrom94,milgrom99,milgrom02,milgrom06,milgrom11,milgrom22a} are time nonlocal.
\section{Some lower-tier phenomena where predictions of  MI and MG may differ materially  \label{lowertier}}
One of the purposes of discussing the specific MI MOND models below in Sec. \ref{models}, is to demonstrate, with concrete examples, some of the potential variety among MOND theories in making secondary predictions. In contradistinction, the primary MOND predictions are those that follow from only the basic tenets of MOND and hence are common to all MOND theories that embody these tenets. Many of these primary predictions and the ways they follow from the basic tenets are discussed in Ref. \cite {milgrom14a}.
\par
But before discussing the specific MI models, and a sample of their predictions, I now described in more general terms some such secondary phenomena and effects, and explain why they might emerge and appear differently in different MOND theories.
\subsection{External-field effect  \label{efegen}}
An important effect, idiosyncratic to MOND, is the external-field effect.  It acts on the internal dynamics of a small, self-gravitating system that is embedded (e.g., falling freely) in the field of a mother system. The external field, even if it is uniform across the subsystem (so no tidal effects are present) makes itself felt in the internal dynamics. The general effect being to suppress MOND effects and bring the MOND predictions closer to those of Newtonian dynamics.
\par
The centrality of the effect was pointed out and exemplified with some specific applications in Ref.  \cite{milgrom83} based on the pristine formulation of MOND, and studied theoretically in detail in the AQUAL theory in Ref. \cite{bm84}.
\par
The EFE is not allowed by the strong equivalence principle, which is obeyed by GR, and there is no explanation of it in standard dynamics with dark matter. So, a robust identification of its presence at low accelerations would constitute a ``smoking gun'' in favour of MOND.
\par
There is a large number of theoretical, phenomenological, and observational studies of this MOND effect in the context of various systems and phenomena.
Some examples of applications are as follows:
Reference \cite{brada00} showed that the EFE by a satellite galaxy can cause a warp in the outer parts of a mother disc galaxy, demonstrating this with the EFE of the  Magellanic clouds on the Milky Way.
Reference \cite{brada00a} described the EFE on dwarf satellites as they move in the field of a mother galaxy, and Ref. \cite{mm13} finds signs for an EFE in the internal dynamics of some satellites of the Andromeda galaxy. Reference \cite{tiret07} discussed the EFE on satellites of elliptical galaxies.
References \cite{muller19,haghi19} discussed the importance of the EFE in ultradiffuse galaxies.
References \cite{haghi16,hees16,wu15,chae20,chae21,stiskalek23,desmond23} showed that the MOND predictions of rotation curved perform somewhat better if one includes the EFE from surrounding structure. Moreover, Refs. \cite{chae20,chae21} found evidence that the small departure of the rotation curves from flatness in the outer parts correlates with their estimates of the external accelerations produced by large-scale structure, in the way predicted by AQUAL/QUMOND.
Reference \cite{asencio22} demonstrated that the Fornax cluster EFE on its dwarf-galaxy members makes them vulnerable to tidal distortion and disruption, in accordance with observations.
The EFE of the Galactic field on star clusters may cause asymmetries in their tidal tails (otherwise hard to explain in the DM paradigm). References \cite{thomas18,kroupa22} discussed and described observations of such asymmetries in globular and open clusters.
\par
The EFE of the Galactic acceleration near the Sun  -- whose strength is $\approx 2\az$ -- greatly quenches MOND effects in the internal dynamics of wide binaries in the solar neighborhood \cite{hernandez12,bps21,chae23,hernandez23,chae23a,hernandez23a,banik23}. Because we are near the MOND-Newtonian transition, the degree of such quenching is sensitive to the exact predicted workings of the effect.
\par
The EFE is also responsible for quenching, practically completely, all MOND effects in experiments on Earth, where the (external) gravitational and kinematical accelerations are many orders of magnitude larger than $\az$.
\par
This extensive body of work attests to the ubiquity and importance of the EFE, making it doubly important that we appreciate the possible variety in its predicted workings in different MOND theories, and  that we recognize which aspects of it are primary and which can differ among MOND theories.
\par
The presence of the effect is quite generic in MOND, in the sense that it is difficult to avoid its appearance in one form or another in any MOND theory \cite{milgrom14a}, but its exact strength and detailed workings can differ substantially among MOND theories.
\par
I will try to make some general and schematic statements, concentrating on purely gravitational systems. Suppose the characteristic MOND acceleration of the external field is $g\_{ex}$ (it can have other characteristics, such as the presence of many variable accelerations and their variability time scales, etc.) and that internal to the system is $g\_{in}$. There is no primary prediction of MOND for the case where $g\_{ex}\ll g\_{in}$ -- such as when considering the effect of the galactic field on the high-acceleration dynamics in the inner Solar System -- not even when $g\_{ex}\sim g\_{in}$.
\par
In situations where $g\_{ex}\gg g\_{in}$, the total acceleration everywhere in the system is $\approx g\_{ex}$, and {\it inasmuch as this is constant}, we can expect the MOND correction over Newtonian behavior to be the same everywhere in the subsystem; namely, that the Newtonian accelerations are multiplied by the same ``discrepancy'' factor everywhere in the system. This would imply roughly that the internal dynamics should be quasi Newtonian with an overall correction factor that can be viewed as a renormalization of G. On dimensional grounds this factor can depend only on $g\_{ex}/\az$ and on additional dimensionless attributes of the system.
So, {\it schematically}, we expect an overall correction, encapsulated in renormalizing $G$ to an effective value
\beq G_e=G\eta(g\_{ex}/\az,\a\_1,\a\_2,...),  \eeqno{renana}
with additional departures from Newtonian dynamics that depend on the dimensionless attributes $\a\_k$.
When $g\_{ex}/\az\rar\infty$, the MOND tenets require that $\eta\rar 1$. In the opposite limit, $g\_{ex}/\az\rar 0$ (but still $g\_{ex}\gg g\_{in}$), scale invariance of the deep-MOND limit dictates that\footnote{The reason for this expectation is that in the Newtonian-like internal dynamics that we have here, internal accelerations scale as $a\sim G_eM/r^2$. Under space-time scaling, $g\_{ex}$, which is a MOND acceleration due to external bodies, transforms as  $g\_{ex}\rar \l^{-1} g\_{ex}$.  We thus have to have $G_e\propto \az/g\_{ex}$, for $a\rar \l\^{-1}a$, as it should ($\az$ appears for dimensional reasons). Another way to see this is to note that scale invariance dictates that only the product $A_0=G\az$ can appear in any deep-MOND-limit prediction, not $G$ and/or $\az$ separately. Thus, we must have $\eta(x,...)\propto 1/x$, for $x\rar 0$.}
\beq \eta=\frac{\az}{g\_{ex}}s(\a\_1,\a\_2,...).    \eeqno{geffa}
AQUAL and QUMOND are underlain by a single interpolating function of a single variable, $\m(x)$. So, inevitably, it is this one function that dictates both rotation curves via $g\_N(r)\approx a\m(a/\az)$, and the above function $\eta$ that enters the prediction of the EFE for dominant external fields $\eta(x,...)=1/\m(x)$.\footnote{In these theories the direction of the external field (encapsulated in some of the $\a\_k$ parameters) is also felt in the internal dynamics -- in a somewhat different way for the two theories \cite{bm84,milgrom10,bz15,chaemil22}.}
\par
But this simplification is not at all a general MOND prediction. {\it The phenomena of rotation curves and the EFE are quite unrelated in MOND, and the rotational acceleration in a disc galaxy, is not the same quantity as the external acceleration that appears in the EFE. There is no reason why the two phenomena should be tightly related in MOND, and underlain by the same formula -- as happens in the special cases of AQUAL/QUMOND.}
So, the function $\eta(x,..)$ may have little to do with the function $1/\m$ that underlies rotation curves in more general MOND theories, as we shall exemplify below.
\par
Specifically, as regards the deep-MOND limit, for example:
The basic tenets do tell us robustly -- making it a primary prediction -- that in the low-acceleration, asymptotic region of rotation curves, the rotational speed is $V(r)\approx\vinf\equiv (MG\az)^{1/4}$ ($M$ is the mass of the galaxy). So the ratio of the MOND acceleration, $a(r)=V^2(r)/r$, and the Newtonian gravitational acceleration there, $\gN=MG/r^2$, is robustly predicted to be $\eta\approx\az/a$ (with the standard normalization of $\az$). However, there is nothing in the basic tenets that tells us that in the prediction of the mass anomaly in subsystems in a dominant, deep-MOND external field, $g\_{ex}$ we also have $\eta\approx\az/g\_{ex}$ -- namely that in Eq. (\ref{geffa}) we have $s=1$.

\subsubsection{Implications of a stronger EFE}
One of the predictions of the specific MI models I discuss below, are that the EFE may be stronger than is predicted by AQUAL/QUMOND. The way this happens in the models might happen more generally: Even in the restricted case whereby the EFE is controlled by the same interpolating function as appears in the predictions of rotation curves, some $\m(x)$, its argument that enters the strength of the EFE need not be $x=g\_{ex}/\az$ itself, but $\a g\_{ex}/\az$, where $\a$ is one of the dimensionless system parameters that can enter the EFE in theories that are underlain by more general interpolating schemes than AQUAL/QUMOND. In the MI models below, $\a>1$ is a function of the ratio of frequencies of the external field variations and the internal ones.
\par
A stronger EFE, as predicted by some MOND theories, stronger even by a factor of a few, may have important implications in cases where $g\_{ex}\not\gg\az$ (when $g\_{ex}\gg \az$ the internal dynamics is approximately Newtonian anyhow). The following are some examples:
\par
In the outskirts of disc galaxies, a stronger EFE may show itself more clearly in causing a more pronounced decline in the outer parts of rotation curve, in departure from the asymptotic flatness predicted for isolated systems, affecting such analyses as in Refs. \cite{haghi16,hees16,wu15,chae20,chae21,stiskalek23,desmond23} .
It can also increase the effectiveness of satellites inducing warps in a mother disc, beyond what is predicted by AQUAL (or QUMOND), as described in Ref. \cite{brada00}.
\par
The field of the Galaxy at the position of the solar system is $\approx 2\az$. The EFE is then expected to quench MOND effects in nearby systems of intrinsic accelerations $\lesssim\az$, such as wide binary stars. Indeed, analyses of Gaia data of such binaries \cite{hernandez12,bps21,chae23,hernandez23,chae23a,hernandez23a,banik23} compare their results with the predictions of AQUAL/QUMOND that give $\eta\approx 1/\m(2)\approx 1.5$. This leads to a quasi-Newtonian behavior, but with velocities elevated by a factor $\eta\^{1/2}\approx 1.2$.
However, if the correct argument value for $\m$ is several times larger, the expected velocity anomaly is even smaller (e.g., only 8\% if $\a=3$).
\par
In MOND, vertical dynamics in disc galaxies cannot be separated from the radial dynamics, because total accelerations are important. The way the radial acceleration -- largely the one that causes the circular rotation in the disc -- enters the vertical dynamics may be viewed as a special case of the EFE.
All analyses of vertical disc dynamics in MOND to date, e.g., in Refs. \cite{bienayme14,loebman14,angus15,angus16,milgrom15,lisanti19}, have adopted the simplistic, single-interpolating-function scheme using either the AQUAL/QUMOND predictions, or even, unjustifiably, assumed the algebraic relation between the Newtonian and MOND accelerations outside the disc. (There is no MOND theory that predicts this; MI predict such an algebraic relation only for exactly circular orbits in the disc's midplane.)
Such analyses would require major modifications in the framework of other MOND theories.
\subsection{Effects of the Galactic field in the Solar System   \label{galasol}}
It has been shown that AQUAL and QUMOND predict an anomaly (departure from standard dynamics) due to the presence of the gravitational field of the Galaxy in which the Solar System is embedded \cite{milgrom09a,novak11}. The anomaly is an added contribution to the solar field that is very small, but may be detectable, even in the inner Solar System, where the acceleration due to the Sun, $g\_\odot(\vr)$, is many orders of magnitude larger than $\az$. This anomaly does not decrease like $1-\m(g\_\odot/\az)$, as some other MOND anomalies, and so it does not become negligible even if $\m(x)$ approach 1 very fast at large $x$.
\par
To understand why this is not a primary MOND prediction, and why it can differ greatly from the predictions of MI, I next explain the origin of this anomaly.
It arises as follows: one solves the AQUAL or QUMOND gravitational field of the Sun with the Galactic field as boundary condition at infinity, and then subtract the latter to get the acceleration relative to the Sun. The MOND gravitational potential, $\f(\vr)$, in such MG theories, can be thought of as the Newtonian potential of a modified source density, $\hat\r=\Delta\f/\fpg$. The excess, $\r\_p\equiv \hat\r-\r$ is called the ``phantom mass density,'' and its Newtonian field constitutes the anomaly.
\par
There is a region around the point -- somewhat different in AQUAL and QUMOND, but several thousand astronomical units away from the Sun -- where the Galactic field cancels that of the Sun, where the gravitational accelerations are $\lesssim\az$, and where AQUAL and QUMOND predict some distribution of phantom matter. This is then felt everywhere in the Solar System as an additive anomaly -- the Newtonian field of the phantom matter -- which is independent of the local value of the acceleration. In the inner solar system, much nearer the Sun than the balance point, this anomaly can be approximated by a quadrupole field, where at a distance $r$ from the Sun, this added acceleration is $\approx q\az(r/\RM)$ (with some quadrupolar angular dependence), where $\RM\equiv (\msun G/\az)^{1/2}\approx 7000$ astronomical units is the ``MOND radius'' of the Sun, and q is a dimensionless prefactor, which is roughly proportional to $1-\m(g\_g/\az)$, with $g\_g\approx 2\az$ the Galactic field at the Sun's position \cite{milgrom09a,novak11}. So, as expected, the value of the anomalous acceleration, at position $\vr$, does not depend on the local value of $g\_\odot$.
The relative correction to the Sun's Newtonian acceleration is $\approx q(g\_\odot/\az)^{-3/2}$, which decreases only as a power $-3/2$ of the ratio $g\_\odot/\az$, no matter how fast MOND approaches Newtonian dynamics at high acceleration, as described by $1-\m(x)\rar 0$. The MOND-to-Newtonian transition occurs when $g\_g/\az$ becomes large, not when $g\_\odot/\az$ becomes large.
\par
Phenomenological consequences and constraints that derive from this anomaly were discussed in Refs. \cite{milgrom09a,novak11, hees16}.
Some studies have, in fact, attributed observed Solar-System anomalies to this effect \cite{paucoklacka16,pauco17,migaszewski23,mathur23}.
\par
In a sense, this anomaly is the special case of the EFE when the external acceleration is much weaker than those in the region of the subsystem we are interested in. But there is an important difference: The main features of the EFE when the external field dominates over all internal accelerations seem generic in MOND, despite some important differences in detail. But the present anomaly is not generic; in some MI formulations it is predicted to essentially not be present.
\par
It has to be realized that the anomaly, predicted by AQUAL and QUMOND,  is the result of idiosyncracies of these theories, specifically, the Poisson-like nature of their field equations. These, like the Poisson equation itself, are elliptic equations, which introduces an element of space-nonlocality. The solution of these equation in a certain region depends on what happens in other regions, in the form of sources, or of dictated boundary conditions.\footnote{For example, in electrostatics the solution depends on the charge density everywhere, and on the presence of conductors that dictate a constant potential on their surface.}
Thus, it is enough that there is some region in space where the MOND gravitational accelerations are small -- such as here, the region around Galaxy-Sun balance point -- to induce an anomalous acceleration everywhere, independent of whether the local total acceleration is very high.
In contradistinction, as has been explained already in Ref. \cite{milgrom09a}, MI theory might make very different predictions on this phenomenon, as indeed will be demonstrated by the specific MI models I describe below.\footnote{A quote from Ref. \cite{milgrom09a}: ``Small, nonrelativistic MOND anomalies in high-acceleration motions may possibly have several
origins. In non-local, modified inertia formulations of MOND, anomalies may appear in the inner
solar system in the motion of bodies that are on highly eccentric trajectories; trajectories that take
them to large distances, where accelerations are low (Milgrom, in preparation). Cases in point are
the long-period comets, and the Pioneer spacecraft. Because of the non-locality of such theories,
the motions in high-acceleration portions of the orbit may be affected by the low acceleration ones,
giving rise to small anomalies. Such MOND effects have been proposed as a possible mechanism for
generating the Pioneer anomaly, without affecting the motions of planets, whose orbits are wholly
in the high acceleration regime (Milgrom 2002, 2005, in preparation).''  \label{comet}}
In MI theories -- such as our MI models below -- the MOND modification depends only on what happens along the trajectory itself. If all accelerations there, $a$, are high, as happens with bodies that live only in the inner Solar System -- MOND approaches Newtonian dynamics as fast as $1-\m(a/\az)$. It does not matter that there are some other regions in space where the accelerations would be small.

\subsection{Additional secondary phenomena}
One important effect, which I mention here only in passing because I cannot offer any concrete statements as to how it works in MI -- but which might be described very differently in MG and MI -- is dynamical friction. The MOND predictions for this effect in AQUAL and QUMOND have been analyzed analytically and numerically, and applied, e.g., in Refs. \cite{ss06,tiret07a,nipoti08,ss09,angus09,vakili17,bilek21,bilek23}.
\par
Gravitational waves are a relativistic phenomena and so goes beyond the basic tenets of MOND, and cannot be thought of as a primary MOND phenomenon. Indeed, existing MG relativistic extensions of MOND show a large variety in what they predict for gravitational waves: the type of modes there are, and their propagation.
\par
Without having any specific result, my only observation at this point is that we can expect very different predictions for gravitational waves in purely MG, relativistic theories, and in MI theories.
Since MG theories modify gravity itself, they necessarily modify the properties of gravitational waves in major ways, generically adding modes to the two general-relativistic propagation modes, modifying their speeds, etc. This may not be necessary in MI theories.
\par
There is at least one class of MG, relativistic extensions of MOND where the tensor mode of gravitational waves propagate in the same way as photons \cite{sz19}.
But this is by no means a generic result; for example, it is not so in the original form of TeVeS.
In MI it is more natural for the path followed by gravitational waves (gravitons) to be the same as that for photons.

\section{A class of Modified-inertia MOND models  \label{models}}
Here I describe, somewhat briefly, the class of MI, MOND models propounded recently in Ref. \cite{milgrom22a}.
I call them ``models'' because, as discussed in Ref. \cite{milgrom22a}, there are some matter-of-principle issues regarding them still to be investigated, for example, the issue of causality. But these models are quite useful as heuristic tools for demonstrating how MI formulations of MOND can differ substantially in their secondary predictions, from the two MG formulations we extensively use today.
These new models are also rather more amenable to analysis than MI models propounded before, such as in Ref. \cite{milgrom94}.
Beyond the summary description of the models, and the discussion of some of their predictions discussed in Ref. \cite{milgrom22a}, I discuss here some new ramifications.
\par
In constructing the models, I do not start from an action principle, but directly from the equations of motion.
\par
The models aim at describing the dynamics of a system of (pointlike) bodies of masses $m\_p$, whose trajectories, $\vr\_p(t)$, the models should dictate. Each body $p$ is subject to {\it Newtonian} forces $\vf\_{pq}(\vr\_p,\vr\_q)$ from other bodies $q$ ($\vf\_{qp}=-\vf\_{pq}$).
So,  body $p$ is subject to the total, time-dependent force
\beq \vF_p(t)=\sum\_{q\not = p}\vf\_{pq}[\vr\_p(t),\vr\_q(t)]   \eeqno{totalforce}
along its trajectory.
Concentrate on one of the bodies; its Newtonian law of motion (omitting its index $p$) is $m\va\_N(t)=\vF(t)$ reads in Fourier space
\beq m \haNo=\hFo.  \eeqno{newfour}
We modify the left-hand, inertia term, writing for the Fourier component of the particle's MOND acceleration, $\hao$,
\beq m\hao\I[\{\hat\vr\},\o,\az]=\hFo,  ~~~~~{\rm or}~~~~~ \hao\I[\{\hat\vr\},\o,\az]=\hat\va\_N(\o).\eeqno{law}
The ``inertia functional'', $\I[\{\hat\vr\},\o,\az]$, is a dimensionless functional of the whole trajectory [designated $\{\hat\vr\}$ to distinguish it from the number $\hro$], and a function of $\o$, and of $\az$.
\par
$\I$ has to conform to the two limits defined in the basic tenets of MOND.
\beq \I~~ \overset{\az\rar 0}{\longrightarrow}~~ 1,  \eeqno{highal}
for the theory to tend to the Newtonian relation (\ref{newfour}) at high accelerations.
And, Ref. \cite{milgrom22a} explains why in the opposite limit, the scale invariance of the deep-MOND limit dictates that\footnote{This follows when we consider a system where the forces are purely gravitational -- what we know of MOND so far all comes from considering gravitational systems. In the context of more general forces, the exact definition of the deep-MOND limit requires further deliberation.}
\beq \I~~ \overset{\az\rar \infty}{\longrightarrow}~~   \A[\{\hat\vr\},\o]/\az,   \eeqno{dmlofI}
where $\A[\{\hat\vr\},\o]$ is a functional of the trajectory, and a function of $\o$, having the dimensions of acceleration.
\par
In Ref. \cite{milgrom94} (Sec. III there), I described some nonlocal, MOND kinetic actions, written in the time domain. The resulting equations of motion, when written in frequency domain do have the form (\ref{law}). So, there are models of this type that are underlaid by an action. But here, I wanted to consider the more general starting point, that of the equations of motion themselves, without necessarily requiring an action origin.
\subsection{Some general properties}
Note that the modified equations of motion (\ref{law}) lead to a time nonlocal theory, since $\{\hat\vr\}$ appearing as an argument in $\I$ requires knowledge of the whole trajectory. Note also that $\hFo$,  or $\hat\va\_N(\o)$ appearing on the right-hand sides in Eq. (\ref{law}), while they are the Newtonian expressions, are calculated for the MOND trajectories (not the Newtonian ones), which are not known a priori.
So the problem has to be solved self consistently.
\subsubsection{Conservation laws}
The requirements for the models to enjoy the standard symmetries: time- and space-translations, rotations, and Galilei invariance, are discuss at length in Ref. \cite{milgrom22a}. I shall not repeat them here, only to say that the specific examples I concentrate on below satisfy these requirements.
These symmetries do not ensure the standard conservation laws without there being an action basis for the theory. Nevertheless, while we still lack such an action for the present models, they do satisfy conservation laws for appropriately defined momentum, energy, and angular momentum. (Remember that modifying inertia always necessitates redefinition of these quantities, as exemplified by special relativity.) These conserved quantities are defined here through their Fourier transform, and are time nonlocal in nature.
We define the momentum, for example, such that it reduces to the standard momentum $m\vv$ in the limit $\az\rar 0$, and that it satisfies $\dot\vP(t)=\vF(t)$, which in Fourier space reads $i\o\hat\vP(\o)=\hFo$. This leads, from Eq. (\ref{law}), to the definition
\beq \hat\vP(\o)=m\hat\vv(\o)\I[\{\hat\vr\},\o,\az].  \eeqno{momentum}
In isolated systems $\sum_p\vF_p=0$; so we have $d(\sum_p\vP_p)/dt=0$.
\par
Similarly, a kinetic energy that satisfies
$dE_k/dt=\vv(t)\cdot\vF(t)$ will give a conserved total energy  when the interbody forces are derivable from potentials, $\f_{pq}(\vr_{pq})$,
since then
\beq \frac{d}{dt}( \sum_p E^k_p+ \sum\_{q<p}\f_{pq})=0.   \eeqno{energy}
This leads to the definition \cite{milgrom22a}
\beq \hat E_k(\o)=\frac{m}{2\pi}\int \frac{\o'}{\o} \hat\vv(\o-\o')\cdot\hat\vv(\o')\I[\{\hat\vr\},\o',\az]d\o',  \eeqno{endef}
which reduces to the standard $E_k=mv^2/2$ for $\az\rar 0$ ($\I\rar 1$).
Similarly, we define the angular momentum as
\beq \hat \vJ(\o)=\frac{m}{2\pi}\int \frac{\o'}{\o} \hat\vr(\o-\o')\times\hat\vv(\o')\I[\{\hat\vr\},\o',\az]d\o', \eeqno{anga}
whose inverse Fourier transform is conserved in systems not subject to an external moment.
\subsubsection{``Center of mass''}
A natural (nonlocal) definition of the center of mass of the many-body system is
\beq \hat\vR(\o)=\frac{\sum_p M_p\hat\vr_p(\o)}{\sum_p M_p},~~~~~~M_p(\o)\equiv m_p\I[\{\hat\vr_p\},\o,\az],   \eeqno{com}
which by the conservation of momentum for isolated systems, satisfies $\ddot \vR=0$, and reduces to the standard definition for $\az\rar 0$.

\subsubsection{The two-body problem in the deep-MOND regime}
As in Newtonian dynamics, the two-body problem in the deep-MOND limit of the present models can be reduced to the problem of a single body in a force field \cite{milgrom22a}, for which the equation of motion in Fourier space is
\beq \bar m\hat\va_{12}(\o)\frac{\A_{12}(\o)}{\az}=\hFo, \eeqno{reduced}
where the reduced mass is given by
\beq \bar m=\frac{m_1 m_2}{(m_1^{1/2}+m_2^{1/2})^2}   \eeqno{redmass}
to be compared with the Newtonian reduced mass $\bar m=m_1 m_2/(m_1+m_2)$.
The quantities with subscript $12$ refer to the relative distances and accelerations, and $\A$ is what defines the deep-MOND limit as in
Eq. (\ref{dmlofI}).
\par
The relative velocity, $V_{12}$, between two masses, $m\_1$ and $m\_2$, on deep-MOND limit {\it circular} orbits around each other is given by\footnote{For circular orbits, there is only one frequency appearing, and $\A_{12}$, which has dimensions of acceleration, can only be proportional to $V^2_{12}/R_{12}$. This is why such a general result can emerge without specifying details of the model. Arbitrary constants that appear are then absorbed into the definition of $\az$, which we always take so that  for a test mass around a larger mass we have $V_{12}^4=MG\az$.}
\beq  V^4_{12}= (q_1^{1/2}+q_2^{1/2})^2MG\az,   \eeqno{circulla}
with $M=m\_1+m\_2$, and $q\_p=m\_p/M$.
This is to be contrasted with the prediction of AQUAL, QUMOND, TRIMOND, and, in fact, the general class of MG theories described in Ref. \cite{milgrom14b}:
\beq V^4_{12}= \frac{4}{9}\left(\frac{1-q\_1\^{3/2}-q\_2\^{3/2}}{q\_1q\_2}\right)^2MG\az .  \eeqno{binarmg}
When $q\_1\ll q\_2$, both expressions give the MOND mass-asymptotic-speed relation (``baryonic Tully-Fisher relation''), $V_{12}^4=MG\az$. This is indeed a primary MOND prediction. But the general-mass-ratio case is not.
For equal masses the coefficient in Eq, (\ref{circulla}) is 2, and in Eq. (\ref{binarmg}) it is $\approx 0.61$, predicting a relative speed $\approx 25$ percent lower for a given $M$.
\par
The result that in the deep-MOND limit $V^4_{12}/MG\az$ has to be of order unity is also a primary prediction that follows from the basic tenets of MOND; but, as exemplified here, the exact value of the ratio is theory dependent, and constitutes a secondary prediction.
\par
Note that relation (\ref{binarmg}) is a special case of a general, deep-MOND, $\s-M$ relation that was shown in Ref. \cite{milgrom14b} to hold in a more general class of MG theories
\beq \s^4=\frac{4}{9}(1-\sum_pq\_p\^{3/2})^{2}MG\az,   \eeqno{gengen}
where $\s$ is the mass-weighted, 3-D, rms velocity in a system made of gravitating point masses, $m\_p$, and $q\_p=m\_p/\sum m\_p$.
As Eq. (\ref{circulla}) already tells us, this specific coefficient in the $\s-M$ relation is not shared by all MOND theories.

\subsection{Subclass of models}
So far, not much was specified about the inertia functional, $\I$, besides its asymptotic forms.
To enable us to discuss possible predictions more concretely, I first limit the discussion to a subclass of models, and then I will further concentrate on specific examples within this subclass.
\par
In the subclass of model, the inertia functional is expressed as  a function of a finite number of variables
\beq  \I[\{\hat\vr\},\o,\az]=\bar\m\left[\frac{\A_1(\o)}{\az},\frac{\A_2(\o)}{\az},...\right],   \eeqno{subclass}
where $\A_i(\o)$ are themselves functionals of the trajectory, and explicit functions of the frequency, with the dimensions of acceleration.
In the limit where all $\A_i\gg \az$ (formally $\az\rar 0$) we should have $\bar\m\approx 1$ to restore Newtonian dynamics.
\par
In the deep-MOND limit, where $\A_i\ll \az$, or, formally, when $\az\rar\infty$, scale invariance of gravitational dynamics -- for which the right-hand side of the equation of motion (\ref{law}) has a scaling dimension $-2$ --  dictates that $\bar\m$ becomes homogeneous of degree 1 in its variables,\footnote{Since the variables $\A_i$ are constructed only from the particle-trajectory degrees of freedom, and since they have dimensions of acceleration, their scaling dimension must be that of kinematic acceleration, namely -1: under space-time scaling, $(t,\vr)\rar\l(t,\vr)$, we have $\A_i\rar \l\^{-1}\A_i$.} namely that in this limit $\bar\m$ has the property that $\bar\m(\l\^{-1} x\_1,\l\^{-1} x\_2,...)=\l\^{-1}\bar\m(x\_1,x\_2,...)$.
This means (seen by taking $\l=x\_1$) that the deep-MOND limit of $\bar\m$ can be written most generally as
\beq \bar\m\overset{\az\rar \infty}{\longrightarrow}\frac{\A_1}{\az}K\left(\frac{\A_2}{\A_1},\frac{\A_3}{\A_1},...\right).  \eeqno{ratiala}
\par
Because different acceleration functionals $\A_i$ can be defined, we see that in this subclass of models the transition from the Newtonian to the deep-MOND limit can be rather more involved than in AQUAL/QUMOND. For example, in the latter the deep-MOND limit behavior is totally fixed, in the former we still have to dictate a function $K$ of several (dimensionless) variables.
This would lead, in effect, to different interpolating functions governing different phenomena, or even different types of trajectories
(e.g., circular ones vs radial ones).
\par
We can also write expression (\ref{subclass}) as
\beq  \I[\{\hat\vr\},\o,\az]=\bar\m\left[\frac{\A_1(\o)}{\az},\B_1(\o),\B_2(\o),...\right],   \eeqno{subclassal}
Where $\B_i$ are dimensionless functionals ($\B_i=\A_{i+1}/\A_1$), and then in the deep-MOND limit $\bar\m(x,y\_1,y\_2,...)\rar xK(y\_1,y\_2,...)$.

\subsubsection{Example  \label{exa}}
In Ref. \cite{milgrom22a}, I discussed requirements on the way $\A_i$ have to depend on $\{\hat\vr\}$ and $\o$, so that the models satisfy various desiderata -- for example, the invariance requirements under time and space translation, and under rotations, and the requirement that they predict a correct center-of-mass motion.
I will not repeat this discussion, but only give an example of models that employ two choices of $\A_i$ that satisfy these requirements:
\beq \A_1(\o)=\frac{1}{2\^{3/2}\pi}\int\_{-\infty}\^{\infty}\t_1\left(\frac{\o'}{\o}\right)[\hat
\vr(\o')\cdot\hat\vr^*(\o')]^{1/2}\o'^2 d\o' , \eeqno{aone}
\beq \A_2(\o)=\frac{1}{2\^{3/2}\pi}\int\_{-\infty}\^{\infty}\t_2\left(\frac{\o'}{\o}\right)|\hat
\vr(\o')\cdot\hat\vr(\o')|^{1/2}\o'^2 d\o'.  \eeqno{atwo}
The equation of motion is thus
\beq \hao \bar\m\left[\frac{\A_1(\o)}{\az},\frac{\A_2(\o)}{\az}\right]=\hat\va\_N(\o),   \eeqno{subclatt}
and in the deep-MOND limit
\beq \bar\m\approx\frac{\A_1}{\az}K(\A_2/\A_1),  \eeqno{ratapo}
with $K(X)$ not dictated by the MOND tenets, since it is scale invariant, and the $\A_1$ prefactor gives $\bar\m$ the correct scaling behavior.
\par
To understand the meaning of the two acceleration functionals, we note that for a single Fourier component
\beq \vr(t)=\frac{1}{\sqrt{2}}(\vr\_0\^{} e^{i\o\_0 t}+\vr\_0^* e^{-i\o\_0 t});~~~~~~~ \vr^2(t)=\vr\_0\^{}\cdot\vr\_0^*+|\vr\_0\^{}\cdot\vr\_0\^{}|\cos{(2\o\_0 t+\varphi)}.  \eeqno{sader}
The amplitude vector, $\vr\_0\^{}=\vr\_R+i\vr\_I$ is a complex-valued vector, and
\beq \vr\_0\^{}\cdot\vr\_0^*=\vr\_R^2+\vr\_I^2,~~~~~~~\vr\_0\^{}\cdot\vr\_0\^{}= \vr\_R^2-\vr\_I^2+2i\vr\_R\cdot\vr\_I .\eeqno{vecaca}
The Fourier component of $\vr(t)$ is
\beq \hat\vr(\o)=[\vr\_0\^{}\d(\o-\o\_0)+\vr\_0^*\d(\o+\o\_0)].  \eeqno{sholer}
We see from the second of Eqs. (\ref{sader}) that $(\hat\vr\cdot\hat\vr^*)^{1/2}$, which appears in the integrand of eq. (\ref{aone}), which defines $\A_1$, is the root-mean-square radius for the component, and the integrand is the root-mean-square acceleration weighted by $\t_1$.
We also see that $|\hat\vr\cdot\hat\vr|^{1/2}$, which appears in the integrand of eq. (\ref{atwo}), which defines $\A_2$, measures the departure from circularity of the trajectory: it vanishes for the circular case (for which $\vr\_R^2=\vr\_I^2$ and $\vr\_R\cdot\vr\_I=0$), and equals $(\hat\vr\cdot\hat\vr^*)^{1/2}$ for the radial case.
So, circular Fourier components do not contribute to $\A_2$.
\par
The forms and normalizations of $\t_i(\z)$ are still free.
With the convenient normalization $\t_1(1)=1$, which we adopt, $\A_1$ equals the acceleration $V^2/R$ for an exactly circular trajectory.
It can be seen that for circular, deep-MOND trajectories around an attracting mass $M$, for which $\A_2=0$, and only $\o'=\o$ contributes in Eq. (\ref{aone}), the theory is easily solved and one gets for the orbital speed
\beq  V^4=MG\az/\t_1(1)K(0). \eeqno{mushrat}
The standard normalization of $\az$ in MOND is then achieved by taking $K(0)=1$.
\par
The normalization of $\t_2$ is degenerate with the dependence of $\bar\m$ on $\A_2$, and we can fix it at $\t_2(1)=1$, as well. This dependence on $\A_2$, and, in particular, the behaviour of the deep-MOND limit $K(X)$, hardly affects the study of rotation curves  (see below), and I cannot offer, at present, constraints on them (if we take these models seriously). To constrain them we need information from test particles on noncircular trajectories; e.g., the comparison of acceleration discrepancies as deduced in the same system from circular trajectories (e.g., rotation curves) and noncircular ones (e.g.,  motions of satellites, such as dwarf-galaxies or globular clusters).
\par
It was also argued in Ref. \cite{milgrom22a}, and explained below in Sec. \ref{commot}, that we have to have
\beq \t_i(\z)\overset{\z\rar \infty}{\longrightarrow} 0,  \eeqno{tetainf}
fast enough. This ensures that  the center-of-mass motion of bodies comprising components at high accelerations -- such as stars in a galaxy -- is still MONDian. This is because then the internal, high-acceleration, but also high-frequency, motions, do not affect the low-frequency, lower-acceleration components of the center-of-mass motion. It makes sense to take $\t_i(\z)$ to be monotonic, in which case they also decrease between $0$ and $1$. So, for instance, $\t_i(0)>1$, which has important consequences for the EFE, which depends on $\t_i(0)$, since typically the characteristic frequencies of the external acceleration are much smaller than those of the internal ones (see below).

\subsection{Phenomenology  \label{phenomenology}}
I now discuss some predictions of the restricted class described in Sec. \ref{exa}.
In Ref. \cite{milgrom22a}, I concentrated, for concreteness's sake, on the even simpler models that involve only $\A_1$ as the acceleration functional appearing as the variable in $\bar\m$ (alluding only in passing to the possibility of defining $\A_2$). Even this was enough to demonstrate that different interpolating functions can appear in different phenomena, because the relevant values of $\t_1(\z)$ may differ when there is more than one underlying frequency.
\par
Here, I emphasize that having more than one acceleration variable enriches the possibilities even further, demonstrating even more forcefully the potential diversity of secondary predictions among MOND formulations.
\par
Below, I discuss some phenomenological consequences of the above model examples, mentioning only in passing those already discussed in Ref. \cite{milgrom22a}, and some new results in more detail.
\subsubsection{Generalities  \label{generalities}}
To be able to present some analytic results of the models, we try to identify situations where the relevant trajectories have only a finite number of discrete Fourier components -- ideally only two major ones. In general, this is not possible. The equation of motion (\ref{law}) is to be solved selfconsistently, with the expression of the solution trajectory, $\vr(t)$, substituted in the expression for the force on the right-hand side. Consider even the simpler case of a single particle moving in a given force field (as contrasted with a many-body problem). If one tries to substitute in $\vF(\vr)$ some trial $\vr(t)$ that has a finite number of Fourier components at frequencies $\o\_k$, the resulting $\hat\vF(\o)$ will, in general, have all the combined frequencies $p\o\_k\pm q\o\_n$, which shows that this $\vr(t)$ cannot be a consistent solution.
\par
But, there are situations, where the physical trajectory can be written, approximately, as a sum of ``separable'' components, each characterized by a single frequency -- or a bunch of neighboring frequencies, which we lump into one for heuristic purposes,
\beq \vr(t)=\sum_k \vr\_k(t), ~~~~~~~~ \vr\_k(t)=\frac{1}{\sqrt{2}}(\bar\vr\_k e\^{i\o\_k t}+\bar\vr\^*\_k e\^{-i\o\_k t}).   \eeqno{mufre}
Furthermore, in such instances, the (Newtonian) force itself separates, to a good approximation, into forces that depend on the separate component trajectories:
\beq \vF[\vr(t)]\approx \sum_k\vF_k[\vr\_k(t)].   \eeqno{separa}
\par
An example is the approximate separation between vertical and radial dynamics in a thin galactic disk. For a particle that does not wonder much in $R$ (ideally, that is on a circular orbit), the $R$ dependence of the force in the perpendicular, $z$, direction becomes irrelevant, and vice versa if the particle does not wonder much in $z$.
We can then write
\beq \vF[\vr(t)]\approx F_R(R)\ve\_R+F_z(z)\ve\_z   \eeqno{separa1}
($\ve\_R$ and $\ve\_z$ are unit vectors in the $R$ and $z$ directions, respectively).
Other examples discussed in Ref. \cite{milgrom22a} are the motion in a triaxial harmonic field where the three principal components are exactly separable, and the motion along each axis is still harmonic (single frequency) but the motion frequencies differ from those appearing in the expression for the potential field: $\f=(1/2)\sum \o\_k\^2 x\_k\^2$.
\par
If such an approximation is justified, we can write separate equations of motion for the different components
\beq m\hat\va\_k(\o\_k)\bar\m\left[\frac{\A_1(\o\_k)}{\az},\frac{\A_2(\o\_k)}{\az}\right]=\hat\vF_k(\o\_k).\eeqno{lawla}
Note, importantly, that while the different components are separated on the right-hand side, they are coupled through $\A_1$ and $\A_2$, which are {\it algebraic functions} of all the $\bar\vr\_k$ and $\o\_k$.
In Eq. (\ref{lawla}), we need the values of $\A_1,~\A_2$ at the spectral frequencies themselves.
For the $n$th component we have, with our normalizations $\t_1(1)=\t_2(1)=1$,
\beq \A_1(\o\_n)=\o\_n\^2(\bar\vr\_n\cdot\bar\vr\_n^*)^{1/2}+\sum_{k\not =n} \o\_k\^2(\bar\vr\_k\cdot\bar\vr\_k^*)^{1/2}\t_1\left(\frac{\o\_k}{\o\_n}\right),~~~~~~~\A_2(\o\_n)=\o\_n\^2|\bar\vr\_n\cdot\bar\vr\_n|^{1/2}+\sum_{k\not =n} \o\_k\^2|\bar\vr\_k\cdot\bar\vr\_k|^{1/2}\t_2\left(\frac{\o\_k}{\o\_n}\right). \eeqno{shiluta}
As I stated above, the functions $\t_1,~\t_2$ have to vanish fast enough for high values of their arguments (``fast enough'' -- in the sense discussed in Sec. \ref{commot} below). This implies that the dynamics at a given frequency $\o\_n$ is affected only little by much higher frequencies, $\o\_k$.
\par
But lower-frequency component do affect the dynamics at $\o\_n$. For example, if $\o\_k\ll\o\_n$, the contribution of its acceleration in $\A_1$
is $\approx \o\_k\^2|\bar\vr\_k|\t(0)\approx a\_k\t(0)$ compared with the acceleration at $\o\_n$ itself $a\_n=\o\_n\^2|\bar\vr\_n|$, which is not enhanced by the factor $\t_1(0)>1$.

\subsubsection{Rotation curves -- exact circular trajectories \label{exactrcs}}
For {\it Exact} circular motion in an axisymmetric field, $\A_2=0$. The trajectory has only one frequency; so only $\t(1)=1$ enters in expression (\ref{shiluta}), which gives $\A_1=\o\^2|\bar\vr|= V^2/R$, and we get the standard ``algebraic relation'' between the Newtonian gravitational acceleration, $\gN$, and MOND kinematic accelerations, $a=V^2/R$:
\beq a\m(a/\az)=\gN,  \eeqno{newmo}
where $\m(x)=\bar\m(x,0)$. With the adopted normalization $K(0)=1$ in Eq. (\ref{ratapo}), we have in deep-MOND limit $\m(x)=x$, and so $\az$ has the standard normalization corresponding to $V^4\_\infty=MG\az$, for the asymptotic rotational speed.
\par
This is the original, basic MOND relation put forth in Ref. \cite{milgrom83}, known as the mass-discrepancy-acceleration relation, or the radial-acceleration relation.
Almost without exception, this is the relation that has been used to analyze rotation curves in MOND, in the form (\ref{newmo}), or its inverted form $a=\gN\n(\gN/\az)$.
\par
This algebraic relation is not only a prediction of the simplistic subclass of heuristic models of Sec. \ref{exa}. It was shown in Ref. \cite{milgrom94} that this is a most general result for any MOND MI theory defined in the restricted sense of Sec. \ref{mondcontext}; {\it but it applies only to exactly circular, constant-speed trajectories.} We shall see below, in Sec. \ref{rcvert}, that it takes corrections when there are other motion components, such as motions perpendicular to the galactic disc.
\par
In theories like AQUAL and QUMOND, observations of rotation curves essentially determine the interpolating function, and with it the full theory, with all its predictions. An important lesson, with more general ramifications, learnt from the present models, is that this need not be the case in other MOND theories. Here, observations of rotation curves for exact circular orbits, do not inform us on the value of $\t_1$ for arguments different from 1 -- which enter, for example the EFE, where the frequencies of the internal and external motions are different (see Sec. \ref{efeefe} below). They also tell us practically nothing about the dependence of the inertia functional $\bar\m$ on $\A_2$, which may enter materially in all phenomena other than rotation curves for exact circular motions.

\subsubsection{Rotation curves -- effects of departure from exact circular motions \label{rcvert}}
The test particles used to measure rotation curves of disc galaxies -- such as stars or neutral gas -- do not generally move on exact circular trajectories. They have also components of radial and vertical motions, which modify the prediction of the radial acceleration, and hence of the rotational speed. Furthermore, different types of test bodies probe different parts of the gravitational field; so the departure from exact circular motion depends, in each galaxy, on the particle type. For example, the neutral gas is generally more concentrated near the midplane of the disc, with smaller acceleration components perpendicular to the disc than stars. And, among the latter, older stellar populations form thicker discs than the younger ones and the cold gas; so they have higher perpendicular velocity dispersions, and they sense, on average, larger perpendicular accelerations.
Also, the $z$ motion of the bodies is not harmonic and each type has its own spectrum.
A full treatment of the problem, even within our simple models, would require much work beyond our scope here.
\par
Here I describe an idealized treatment of the problem to get an idea of the possible effects, and of how different attributes enter into it. Assume that the test particles have just two components of motion: an exact circular one -- neglecting radial motions -- with frequency $\o=V/R$, and a component of motion in the perpendicular, $z$, direction, which I approximate by a harmonic linear motion with a single frequency, $\o_z$, with an amplitude $\bar\vr\_k=z\_0\ve\_z$, as defined in Eq. (\ref{mufre}). We then ask how the presence of the $z$ motion affects the rotational speed of that particle.
\par
We employ Eq. (\ref{shiluta}) with the circular component being $n$ and the $z$ component being $k$ in the sums. The former, being circular, does not contribute to $\A_2$, and the second, being linear, enters with the same acceleration in both terms. Thus we have
\beq \A_1(\o)=\frac{V^2}{R}+\o_z\^2z\_0\t_1\left(\frac{\o_z}{\o}\right),~~~~~~~\A_2(\o)= \o_z\^2z\_0\t_2\left(\frac{\o_z}{\o}\right). \eeqno{shilush}
The effect of this correction, is to replace Eq. (\ref{newmo}) with
\beq \frac{V^2}{R}\bar\m(\A_1/\az,\A_2/\az)=\gN.  \eeqno{newmota}
In most of the galactic disc, the correction is expected to be small, because, except near the center, $\o_z\^2z\_0 \ll V^2/R$, and because $\o_z>>\o$, and $\t_i(\z)$ has to decrease from its value of $1$ at $\z=1$. But, as we approach the center, the ratio of the mean $z$ acceleration to the radial one increases, and the ratio of frequencies may decrease; so the predicted correction to the rotation curve may become more important. Inasmuch as the correction is small and we can expand relation (\ref{newmota}) in it, we can write
\beq \frac{V^2}{R}\m\left(\frac{V^2}{R\az}\right)\left\{1+\frac{a\_z}{a\_R}\hat\m\left[\t_1\left(\frac{\o_z}{\o}\right)+
\t_2\left(\frac{\o_z}{\o}\right)\tilde\m\right]\right\}=\gN,  \eeqno{newmochic1}
where $ a\_z=\o_z\^2z\_0,~~~~a\_R=\o\^2R,$ and
\beq \hat\m\equiv \frac{d\ln\m(x)}{d\ln x}=  \frac{\partial\ln\bar\m(x,0)}{\partial \ln x},~~~~~~~~\tilde \m\equiv\frac{\bar\m_{x\_2}(a\_R/\az,0)}{\bar\m_{x\_1}(a\_R/\az,0)}.   \eeqno{juster}
($\bar\m_{x\_i}$ is the partial derivative of $\bar\m$ with respect to $x\_i$.)
\par
We see that even in the simplified, heuristic models we employ here, the  departure of the MOND prediction from that for exact circular motion Eq. (\ref{newmo}) can depend in complicated ways on the local acceleration components and the frequency ratio.
\par
As already noted, we are not able to specify $\tilde\m$. But we know from the successes of MOND phenomenology that $\hat\m$ is positive (and equals 1 in the deep-MOND regime). Thus, inasmuch as $\tilde\m$ is not negative and large enough to give a negative value to the square bracket in Eq. (\ref{newmochic1}), the curly bracket in this equation is larger than 1. And since $x\m(x)$ is increasing, the correction due to $z$ motion {\it reduces} the value of the predicted rotational speed relative to that predicted by the algebraic relation.
\par
To get some idea of the possible corrections, we can hark back to the analysis of
Ref. \cite{brada95}. As us here, they tried to see how a correction to the algebraic relation that accounts in some way for the contribution from $z$ accelerations affects the MOND prediction, and how the corrected prediction compares with the prediction of AQUAL (the touchstone for MOND at the time).
They did this by applying the algebraic relation between the MOND and Newtonian accelerations just outside the disc (marked with a $+$ superscript), instead of in the midplane, namely, they used the relation
\beq  \va\^+\m(|\va\^+|/\az)=\vg\^+\_N.  \eeqno{calkaw}
We need the radial component of this equations for the rotation curves. (Unlike the vertical component, the radial accelerations do not vary across the thickness of a thin disc.) This gives
\beq V^2/R=\gN/\m(a\^+/\az),  \eeqno{alala}
where $\gN$ is the radial Newtonian, gravitational acceleration, and $a\^+\equiv|\va|\^+$ represents the total MOND acceleration just outside the disc, calculated from the Newtonian total acceleration just outside the disc, $\gN\^+=[\gN\^2+(\tpg\S)^2]^{1/2}$,
from $\m(a\^+/\az)a\^+=\gN\^+$. ($\S$ is the disc surface density at position $R$.)
So the correction to the algebraic relation is in modifying the argument of the interpolating function to the total acceleration outside the disc instead of the one in the midplane.
\par
\begin{figure}
  \centering
 \includegraphics[width=.6\columnwidth]{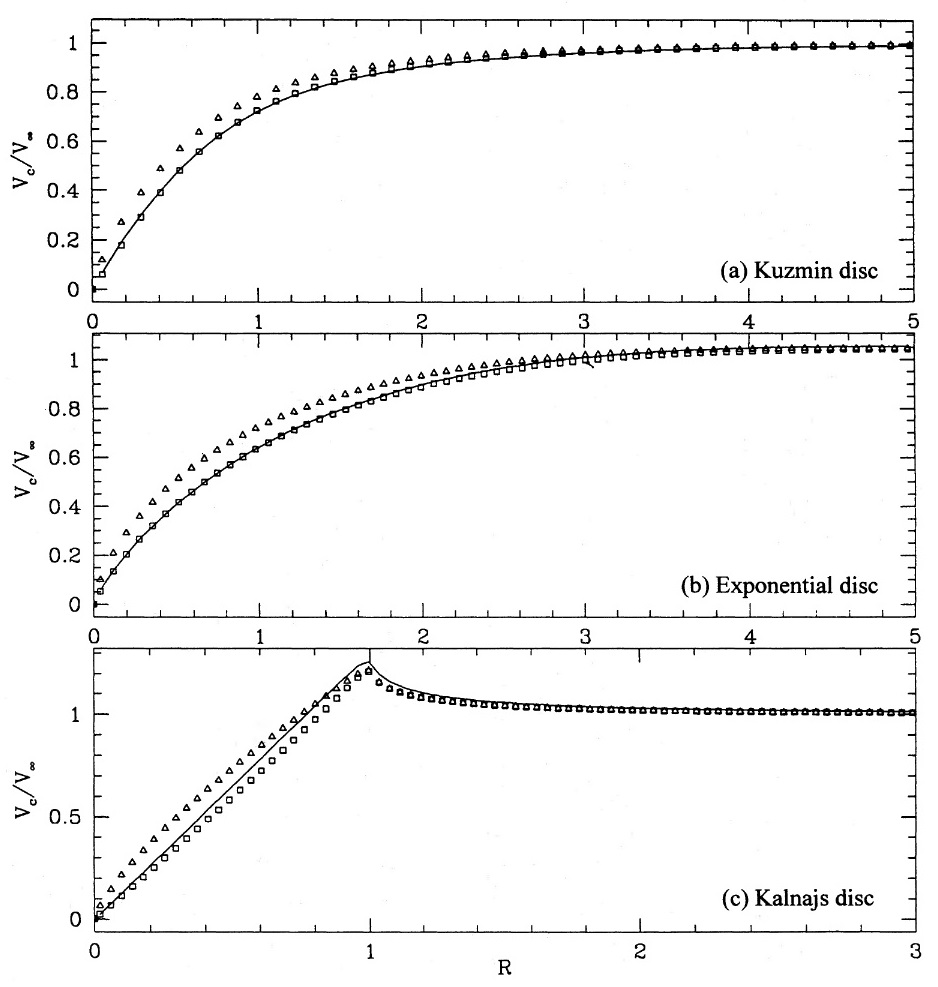}
  \caption{Comparison of the rotation curves predicted by AQUAL (solid line) with the prediction of MI for exactly circular trajectories in the midplane of the disc (the algebraic relations, triangles), and with and algebraic relation given in Eq. (\ref{alala}) (squares), for three pure-disc galaxy models, as indicated. From Ref. \cite{brada95}.}\label{brada}
\end{figure}

This procedure does not account for the effect of the $z$-accelerations on the predicted rotation curves in exactly the way dictated by our MI models, but it does show similarly how the correction is small at large radii, becomes important at small radii, and brings the predicted MI velocities closer to those predicted by AQUAL. All these are shown in Fig. \ref{brada} for three disc models.
\par
Another implication of the necessity of the above corrections it to reduce the predicted value of the $Q$ parameter defined in Ref. \cite{milgrom12} (and studied in Ref. \cite{pl20}) for MI relative to the value for MG, perhaps reducing the difference between them and diminishing the value of this parameter as a discriminant between theories.
\par
Note also that the corrections of the type we discuss above are trajectory dependent, and so may lead to somewhat different predicted rotational speeds for different types of test particles in a disc galaxy. For example, older discs in galaxies -- which are, by and large, thicker and have higher velocity dispersions than the younger stellar discs -- may be predicted to have lower rotational speeds than the thinner, younger components. Indeed, Ref. \cite{vieira22} find that, near the Sun, the Milky Way's thick disc rotates somewhat slower than the thin disc (by about 5 percent, but with high significance). But this could be due to reasons other than those we discuss here (e.g., a higher asymmetric-drift contribution).
\par
The correction discussed here it not a function of only the (Newtonian) acceleration, it depends also on other parameters that vary from point to point, and may even depend on the test particles studied. It is thus predicted to produce some scatter in the observed mass-discrepancy-acceleration ($a-\gN$) relation, on top of the systematic reduction in predicted $a$ values. This may account for the small ``intrinsic'' scatter deduced in Ref. \cite{desmond23}.
\par
We note in this connection that the rotation-curve analysis of Ref. \cite{chae22}, which claimed to have found preference for AQUAL over MI, used for the MI MOND prediction the exact algebraic relation (\ref{newmo}). This analysis thus did not test MI MOND, and would require reevaluation in light of the above conclusions.

\subsubsection{Vertical dynamics in a (galactic) disc}
To get an idea of how the radial acceleration in a disc galaxy affects the perpendicular dynamics, I employ the same idealized treatment as in the previous section. Use Eq. (\ref{shiluta}) with the $z$ component labeled $n$, and the circular component labeled $k$ in the sums. Since the latter is idealized as exactly circular, it does not contribute to $\A_2$. We then have
\beq \A_1(\o_z)=\o_z\^2z\_0+\o\^2R\t_1\left(\frac{\o}{\o_z}\right),~~~~~~~\A_2(\o_z)= \o_z\^2z\_0. \eeqno{shiluba}
By Eq. (\ref{lawla}), the ratio of the Newtonian to the MONDian Fourier components of the accelerations is then
\beq  \frac{\hat g\_N\^z(\o_z)}{\hat a\_z(\o_z)}\approx   \bar\m\left[\frac{\A_1(\o_z)}{\az},\frac{\A_2(\o_z)}{\az}\right].   \eeqno{luoper}
For example, in cases where we can neglect the $z$ acceleration compared with the circular one, $\o_z\^2z\_0\ll \t_1(\o/\o_z)\o\^2R$,
we have
\beq \frac{\hat g\_N\^z(\o_z)}{\hat a\_z(\o_z)}\approx  \m\left[\t_1\left(\frac{\o}{\o_z}\right)\frac{\o\^2R}{\az}\right]   \eeqno{luokal}
[$\m(x)\equiv\bar\m(x,0)$].
Note, importantly, that this is {\it not} the same factor that gives the rotation-curve anomaly at $R$, which is $\m(\o\^2R/\az)$, without the factor $\t_1$ that appears in the argument of expression (\ref{luokal}).
If, e.g., $\o\ll\o_z$, as is many times the case -- for example in the solar neighborhood for most disc components -- then this factor can be approximated by $\t_1(0)$. Since $\t_1$ is assumed to be decreasing and $\t_1(1)=1$, $\t_1(0)$ could be of the order of a few.
\par
The presence of this factor in the argument could greatly change the MOND predictions of $z$ dynamics by th epresent models, compared with those of AQUAL/QUMOND.
\subsection{A system falling in an external field}
Consider the motion of a small composite body, made of gravitating constituents, in the field of a larger body. For example, a star, a gas cloud, or a star cluster falling in the field of a galaxy, or a galaxy falling in the field of a galaxy cluster. Two crucial questions that arise are: (a) How do the internal accelerations of the constituents affect the center-of-mass motion of the whole system? (b) How does the external field affect the internal dynamics?
Both questions have clear-cut and satisfactory answers within existing MG theories such as AQUAL, QUMOND, and TRIMOND.
For MI, I already discussed in Sec. \ref{efegen} generalities of the answer to the second question, which go under the term EFE, and I'll discuss the specific predictions of our model examples below in Sec. \ref{efeefe}. But first I discuss the first question within these models.
\subsubsection{Center-of-mass motion  \label{commot}}
Again, to be able to derive concrete predictions, and treat the body in question as a well-defined entity, with a well-defined overall motion,
I assume that the body is of characteristic size, $r$, much smaller than the scale, $R$, that characterizes its ``external'' trajectory.
So the motions of each constituent can be clearly separated to ``external'' Fourier components, of frequencies collectively designated $\Omega$, and amplitudes, collectively designated $R$, and ``intrinsic'' Fourier components with analogous characteristics $\o$ and $r$. The characteristic external accelerations are then $\sim\Omega^2R$, and the internal ones are $\sim \o\^2r$.
\par
Observations tell us that the overall motion of such bodies obeys MOND in a way that is oblivious to the internal dynamics; e.g., stars of all types, gas clouds, etc. give roughly the same rotation curve, even though they have very different internal structures. Such a behavior is also in line with the weak equivalence principle (universality of free fall) as applied to such composite bodies -- which we wish to retain. Indeed, observations tell us that it holds, at least approximately, also in the MOND regime.
\par
To satisfy this requirement, we need that all constituents share, to a good enough approximation, the same ``external'' components of motion, and that these external components obey the MOND equation appropriate for their own acceleration. A sufficient condition for this is that we can neglect in Eq. (\ref{shiluta}) all the contributions of the internal motions to the ``inertia functional'' $\bar\m$ of the ``external'' motions.
\par
Consider then expressions (\ref{shiluta}) for the ``external'' components. For these, all the contributions of the internal motions appear as terms in the sum over $k$; so they all appear with a factor $\t_i$. Thus their contribution to $\A_i$ relative to the main term, which is of order $\Omega^2 R$, is
\beq \sim \frac{\o\^2r}{\Omega^2R}\t_i\left(\frac{\o}{\Omega}\right).   \eeqno{ooo}
If $\o\^2r\ll\Omega^2R$ the internal accelerations are immaterial for the external ones. If this is not so, then $\o\gg\Omega$, because $R\gg r$.
Thus, even if $\o\^2r\gg\Omega^2 R$, the relative contributions are very small if $\t_i(\z)$ decrease fast enough with $\z$. For example, if $\t_i(\z)\sim \z\^{-2}$ for $\z\gg 1$, expression (\ref{ooo}) is $\sim r/R\ll 1$; in fact, it then becomes of the order of the tidal effects, which we neglect anyhow.
\par
Of course, $\t_i$ could decrease even faster (such as exponentially) which would suppress even more the effect of the internal on the external motions.

\subsubsection{The external-field effect \label{efeefe}}
Consider again the simplified picture where the subsystem falls in the external field on a trajectory whose frequencies are $\Omega_m$ and the corresponding (complex) amplitude vectors are $\bar\vR_m$. The internal motions have frequencies $\o\_k$ with amplitudes $\bar\vr\_k$.
In most past and foreseeable applications, we have $\Omega_m/\o\_k\ll 1$ for all $m$ and $k$, which I assume, and so put these ratios to be zero, without losing much accuracy. I also assume that $|\bar\vR_m|\gg|\bar\vr\_k|$, so the finite size of the subsystem can be neglected, and all constituents can be assumed to have the same ``external'' motion.
\par
If either of these strong inequalities do not hold, we cannot justify the decoupling of motions that underlie the equations of motion (\ref{lawla}), and a rather more complicated treatment would be necessary.
\par
We are now interested in the internal dynamics, and the way they are affected by the external motions. One
first solves for the external motion, which is oblivious to the internal dynamics under our assumptions, as shown above in Sec. \ref{commot}.
Then, the equations of motions are applied to the $\o\_n$ components, with the relevant arguments of $\bar\m$ calculated from Eq. (\ref{shiluta})
$$ \A_1(\o\_n)=\o\_n\^2(\bar\vr\_n\cdot\bar\vr\_n^*)^{1/2}+ \sum_{k\not = n}\t_1\left(\frac{\o\_k}{\o\_n}\right)\o\_k\^2(\bar\vr\_k\cdot\bar\vr\_k^*)^{1/2} +\t_1(0)\sum_m\Omega_m^2(\bar\vR_m\cdot\bar\vR_m^*)^{1/2},$$
 \beq \A_2(\o\_n)=\o\_n\^2|\bar\vr\_n\cdot\bar\vr\_n|^{1/2}+  \sum_{k\not = n}\t_2\left(\frac{\o\_k}{\o\_n}\right)\o\_k\^2|\bar\vr\_k\cdot\bar\vr\_k|^{1/2} + \t_2(0)\sum_m\Omega_m^2|\bar\vR_m\cdot\bar\vR_m|^{1/2}. \eeqno{shilbis}
Thus, the external motion enters as a constant (independent of $\o\_n$), additive contribution to the $\A_i$. Note that I have not assumed that the intrinsic frequencies decouple among themselves -- only from the external motions -- in the sense of Eq. (\ref{separa}). So the values of $\A_i$ from Eq. (\ref{shilbis}) are to be used in the general equation of motion, Eq. (\ref{subclatt}).
\par
The result is rather simple when the external accelerations are much dominant over the internal ones, $\Omega_m^2|\bar \vR_m|\gg\o\_k^2|\bar r\_k|$ (in addition to the other strong inequalities).\footnote{$\t_i$ are decreasing so are maximal at $0$.}
Then
\beq \A_1(\o)\approx \A_1\^{ex}\equiv \t_1(0)\sum_m\Omega_m^2(\bar\vR_m\cdot\bar\vR_m^*)^{1/2},~~~~~~~\A_2(\o)\approx \A_2\^{ex}\equiv \t_2(0)\sum_m\Omega_m^2|\bar\vR_m\cdot\bar\vR_m|^{1/2} \eeqno{shilkuk}
{\it are the same for all intrinsic frequencies}, and the equations of motion (\ref{subclatt}) for the internal dynamics are
\beq \hao \bar\m\_{ex}=\hat\va\_N(\o),\eeqno{lawlade}
where
\beq \bar\m\_{ex}\equiv \bar\m\left[\frac{\A_1^{ex}}{\az},\frac{\A_2^{ex}}{\az}\right]   \eeqno{makaty}
is a constant.
\par
We see that in this case the system obeys, intrinsically, the equations of motion of Newtonian dynamics, $m\bar\m\_{ex}\hao =\hat\vF(\o)$  ($\hat\vF$ is the Newtonian force fourier component), but with all the {\it inertial} masses, $m$, of its constituents, {\it but not the gravitational masses,} renormalized, by a constant factor, to smaller values $m\bar\m\_{ex}$. Specifically, in a gravitating system, this has the same effect as replacing $G$ with the larger, effective $G_e=G/\bar\m\_{ex}$.
\par
In more detail, it can be seen that starting from a Newtonian solution for the trajectories of the constituents $\vr\^N\_p(t)$ gotten by dictating the initial conditions $\vr\^N\_p(0)$, $\vv\^N\_p(0)$, the MOND solution, for the initial conditions $\vr\_p(0)=\vr\^N\_p(0)$, $\vv\_p(0)=\l\vv\^N\_p(0)$, is $\vr\_p(t)=\vr\^N\_p(\l t)$, where $\l=1/\sqrt{\bar\m\_{ex}}$. Thus the orbits are the same as the Newtonian ones, but the particles run along them at speeds $1/\sqrt{\bar\m\_{ex}}$ times higher, $\vv\_p(t)=\l\vv\^N\_p(\l t)$, and accelerations
$1/\bar\m\_{ex}$ times higher, $\va\_p(t)=\l\^2\va\^N\_p(\l t)$.

\par
Compare this result with the analog one (for a purely gravitational system) in AQUAL and QUMOND: A subsystem of characteristic size, frequency, and accelerations, $r$, $\o$, and $g\_{in}\sim\o\^2r$, respectively, in an external acceleration field, $\vg\_{ex}$, that varies on typical length scales $R$, and on time scales $1/\Omega$, and whose momentary value much dominates over the internal ones, $|\vg\_{ex}|\gg g\_{in}$. As above, we also assume here that $r\ll R$, so the external field is approximately constant across the subsystem, and that $\Omega\ll\o$, so the external field can be assumed time independent (over the intrinsic dynamical times), and if it does change, this causes adiabatic changes in the internal dynamics.\footnote{Reference \cite{brada00a} showed that the conditions of nonadiabaticity and substantial tidal effects occur roughly simultaneously for a dwarf system on an elongated trajectory in the field of a mother system.}
Then, AQUAL and QUMOND both predict that the internal dynamics is ``Newtonian like'', but with an effective $G_e=G/\m(g\_{ex}/\az)$, where $\m$ is the interpolating function of these theories.\footnote{In QUMOND $\m(x)$ is uniquely defined by the `inverse' interpolating function, $\n(y)$.} In addition, there appear some distortion effects of order unity, that depend also on the direction of $\vg\_{ex}$ (and differ somewhat between QUMOND and AQUAL) \cite{bm84,milgrom10,bz15,chaemil22}.
\par
These predictions look similar in some sense to the ones above -- encapsulated in Eqs. (\ref{lawlade}-\ref{makaty}) --  that follow from our model MI theories: The intrinsic dynamics is ``Newtonian like'', with a renormalized value of $G_e$ that is Larger than $G$ and is given by $G/\m^*$, where $\m^*$ is some constant that is the value of some interpolation function with arguments that depend only on characteristics of the external acceleration.
\par
But there are also important differences between the predictions as follows: (a) In the MG theories, the effect depends only on the momentary value of $\vg\_{ex}$, and so it changes along the external trajectory. In the MI models, the effect depends on global characteristics of the external orbit, {\it and is the same everywhere along it}. (b) The MG theories predict also some (order unity) dependence on the direction of $g\_{ex}$; not so in the MI predictions. (c) The renormalizing factor $\m^*=\m(g\_{ex}/\az)$ in the MG predictions depends only on one number, $g\_{ex}$; this $\m^*$ is fully determined by what rotation-curve analysis informs us (since it determines the interpolating function). In the MI prediction there appears $\m^*$ that is a function of two variables $\m^*=\bar\m(x_1,x_2)$. But rotation curves tell us nothing about the dependence on $x_2$; and the variable $x_1$ involves the factor $\t_1(0)$, which does not appear in predictions of rotation curves.

\subsubsection{Galaxy effect on the inner Solar System \label{GalaxxyMI}}
I discussed in Sec. \ref{galasol} a secondary prediction of AQUAL/QUMOND, of a possible small MOND anomaly due to the galactic acceleration. This might show up in the motion of bodies in the inner Solar System -- such as planets -- even though their acceleration due to the Sun is many orders of magnitude larger than $\az$ \cite{milgrom09a,novak11,hees16}. I explained why this prediction is idiosyncratic to AQUAL/QUMOND (and possibly to similar MG theories, such as TRIMOND). I also explained why MI theories might predict very differently, which I now demonstrate in the context of our heuristic MI models.
\par
Consider a test particle (e.g., a planet) that moves around the Sun, and stays always at distances from the Sun where its acceleration is $g\_\odot\gg\az$ (unlike, e.g., very-long-period comets whose orbit may reach beyond the MOND radius of the Sun; see footnote \ref{comet} above). The trajectory of the Sun itself in the Galaxy, $\vR(t)$, is assumed oblivious to the effects of the planet. It is thus solved for first, and then it is a given of our problem.
As in the MG case, we take the limit of planetary orbital size much smaller than $|\vR(t)|$; so the Sun and the planet are subject to the same galactic field.\footnote{The real problem, including the effects of other planets, etc. is rather more complicated. But for the comparison between MG and MI theories this simplified description suffices.}
\par
We can then describe the trajectory of the planet as $\vR(t)+\vr\_N(t)+\vr\_a(t)$, where $\vr\_N(t)$ is the Newtonian trajectory -- calculated for some initial conditions -- and $\vr\_a(t)$ is a possible small correction (the anomaly) that might appear in a MOND calculation that includes the galactic field, and which we are after. In principle, there may be a small MOND anomaly even without the presence of the Galactic field, and even at the high accelerations, $g\_\odot$, in the inner solar system, because the Newtonian limit is not yet fully reached even at such high acceleration. In theories such as AQUAL/QUMOND, this can be expressed as the fact that $1-\m(g\_\odot/\az)\not = 0$. This type of anomaly has been considered and constrained , e.g., in Refs. \cite{milgrom83,milgrom09a,sj06}.
This is not the anomaly we are after, and it can be easily put aside and ignored if MOND approaches standard dynamics fast enough at high accelerations. For example, with the $\n$ interpolating function, $\n(y)=1/(1-e^{-y^{1/2}})$ that is now often used for rotation-curve analysis, we have for the Earth $\n-1\approx e^{-10^4}$, totally negligible.
\par
As explained in Sec. \ref{galasol}, the anomaly in AQUAL/QUMOND does not disappear even if we put $1-\m(g\_\odot/\az)= 0$. But, in our MI models, we can use Eq. (\ref{shilbis}) to see that the exact Newtonian solution {\it without an anomaly} applies here since the values of $\A_i$ calculated for it are many orders of magnitude larger than $\az$ in the inner solar system. (For the Newtonian solution, the first terms in Eq. (\ref{shilbis}) are very large.) The galactic field enters only as a small addition to the Newtonian contribution in the arguments of $\bar\m$. So, unlike the AQUAL/QUMOND anomaly, here the anomaly does disappear if the theory tends to Newtonian dynamics fast enough, and so $1-\m(g\_\odot/\az)$ is negligible.

\section{Discussion  \label{discussion}}
 I have tried to bring home the following main points:
\par
(a) It is quite worthwhile, and highly desirable, to explore MOND theories beyond the restricted class of modified-gravity theories, which have been studied almost exclusively, so far.
\par
(b) Roughly speaking, predictions made by MOND theories can be divided into primary ones, those that follow, within a margin of minor latitude, from only the basic tenets of MOND, and secondary ones, which do not. The former are then shared by all MOND theories (defined as those that embody the basic tenets), while the latter can differ greatly among MOND theories, and are much better suited for distinguishing observationally between theories. I thus also warn against viewing predictions of the two workhorse MOND MG theories, AQUAL and QUMOND, as absolute predictions of MOND.
\par
(c) In particular, it is advisable to explore ``modified-inertia'' interpretation of MOND, despite the fact that they seem more drastic modifications of standard dynamics, and perhaps more difficult to devise and explore. Effective appearances of modified inertia, or emergent inertia, are rife in standard physics. And, this avenue seems to me very promising for MOND as well, which, as far as it now appears, is also an effective theory -- an approximation of a more fundamental theory -- a FUNDAMOND. This assessment of MOND as we now know it being an effective theory is based on the fact that all present MOND theories still introduce by hand an interpolation scheme between the standard dynamics (Newtonian and general relativistic) and deep-MOND dynamics. In a FUNDAMOND, we expect that various results would
have a standard-dynamics limit at high accelerations, and a deep-MOND limit at low ones, but not that this should be described by some universal interpolating function that is introduced by hand at the fundamental level of the theory. The proximity of the MOND acceleration constant, $\az$, to accelerations of cosmological significance may also hint at a FUNDAMOND in which such a ``coincidence'' would appear naturally \cite{milgrom99,milgrom20a}.
\par
(d) Modified-inertia, MOND theories are not only worthwhile exploring because they may turn out to be the correct way to FUNDAMOND. In being distant in concept from MG theories, they also demonstrate clearly how MOND theories can differ greatly in predicting the dynamics of secondary phenomena.
This can be understood on general grounds, and demonstrated with specific examples.
\par
To demonstrate and exemplify such possible variety in secondary predictions, I described briefly the class of MI models detailed in Ref. \cite{milgrom22a}, and derive some of their predictions for primary and secondary phenomena.
I stress that I do not propound these models as full-fledged, MI, MOND theories.
There are still unresolved matter-of-principle aspects of these models, and it is also not clear how to apply them in all circumstances, as explained in more detail in Ref. \cite{milgrom22a}.


\begin{thebibliography}{}
\bibitem{milgrom83}M. Milgrom, 1983, A modification of the Newtonian dynamics as a possible alternative to the hidden mass hypothesis. Astrophys. J. 270, 365
\bibitem{fm12}B. Famaey and S.S. McGaugh, 2012,  Modified Newtonian Dynamics (MOND): Observational Phenomenology and Relativistic Extensions. Living Rev.  Relativ. 15, 10
\bibitem{milgrom14}M. Milgrom, 2014, continually updated, The MOND paradigm of modified dynamics. Scholarpedia, 9, 31410
\bibitem{milgrom20}M. Milgrom, 2020, MOND vs. dark matter in light of historical parallels. Stud. Hist. Philos. Mod. Phys. 71, 170
\bibitem{mcgaugh20}S. McGaugh, 2020, Predictions and Outcomes for the Dynamics of Rotating Galaxies. Galaxies 8, 35
\bibitem{merritt20}D. Merritt, 2020, {\it A Philosophical Approach to MOND: Assessing the Milgromian Research Program in Cosmology}, Cambridge University Press
\bibitem{bz22}I. Banik and HS. Zhao, 2022, From galactic bars to the Hubble tension --
weighing up the astrophysical evidence for Milgromian gravity. Symmetry 14, 1331
\bibitem{mm13}S. McGaugh and M. Milgrom, 2013, Andromeda Dwarfs in Light of MOND. II. Testing Prior Predictions. Astrophys. J. 775, 139
\bibitem{milgrom19}M. Milgrom, 2019, MOND in galaxy groups: A superior sample.  Phys. Rev. D 99, 044041
\bibitem{milgrom09c}M. Milgrom, 2009, The central surface density of `dark haloes' predicted by MOND. Mon. Not. R. Astron. Soc. 398, 1023
\bibitem{lelli16}F. Lelli, et al. 2016,  The Relation between Stellar and Dynamical Surface Densities in the Central Regions of Disk Galaxies. Astrophys. J. Lett. 827, 19L
\bibitem{milgrom16}M. Milgrom, 2016,  Universal Modified Newtonian Dynamics Relation between the Baryonic and "Dynamical" Central Surface Densities of Disc Galaxies. Phys. Rev. Lett. 117, 141101
\bibitem{brada99}R. Brada and M. Milgrom, 1999, The Modified Newtonian Dynamics Predicts an Absolute Maximum to the Acceleration Produced by ``Dark Halos''. Astrophys. J. Lett. 512, L17
\bibitem{milgrom89}M. Milgrom, 1989, On Stability of Galactic Disks in the Modified Dynamics and the Distribution of Their Mean Surface-Brightness. Astrophys. J. 338, 121
\bibitem{brada99a}R. Brada and M. Milgrom, 1999, Stability of Disk Galaxies in the Modified Dynamics. Astrophys. J. 519, 590
\bibitem{banik18}I. Banik, HS. Zhao, and M.Milgrom, Toomre stability of disk galaxies in quasi-linear MOND, arXiv:1808.10545
\bibitem{milgrom14a}M. Milgrom, 2014, MOND laws of galactic dynamics. Mon. Not. R. Astron. Soc. 437, 2531

\bibitem{brada95}R. Brada and M. Milgrom, 1995, Exact solutions and approximations of MOND fields of disk galaxies. Mon. Not. R. Astron. Soc. 276, 453
\bibitem{milgrom12}M. Milgrom, 2012, Global Deep-MOND Parameter as a Theory Discriminant. Phys. Rev. Lett. 109, 251103
\bibitem{pl20}J. Petersen and F. Lelli, 2020 A first attempt to differentiate between modified gravity and modified inertia with galaxy rotation curves.  Astron. Astrophys. 636, A56
\bibitem{brown18}K. Brown, R. Abraham, L. Kell, and H. Mathur, 2018,
The radial acceleration relation and a magnetostatic analogy in quasilinear MOND. New J. Phys. 20, 063042
\bibitem{chae22}K.H. Chae, 2022, Distinguishing Dark Matter, Modified Gravity, and Modified Inertia with the Inner and Outer Parts of Galactic Rotation Curves. Astrophys. J. 941, 55
\bibitem{bm84}J. Bekenstein and M. Milgrom, 1984, Does the missing mass problem signal the breakdown of Newtonian gravity? Astrophys. J. 286, 7
\bibitem{sanders97}R.H. Sanders, 1997, A Stratified Framework for Scalar-Tensor Theories of Modified Dynamics. Astrophys. J. 480, 492
\bibitem{soussa03}M.E. Soussa and R.P. Woodard, 2003, A nonlocal metric formulation of MOND. Class. Quant. Grav. 20, 2737
\bibitem{bekenstein04}J.D. Bekenstein, 2004, Relativistic gravitation theory for the modified Newtonian dynamics paradigm. Phys. Rev. D 70, 083509
\bibitem{zfs07}T.G. Zlosnik, P.G. Ferreira, and G.D. Starkman, 2007, Modifying gravity with the aether: An alternative to dark matter. Phys. Rev. D 75, 044017
\bibitem{milgrom09b}M. Milgrom, 2009, Bimetric MOND gravity. Phys. Rev. D 80, 123536
\bibitem{deffayet14}C. Deffayet, G. Esposito-Farese, and R.P. Woodard, 2014, Field equations and cosmology for a class of nonlocal metric models of MOND. Phys. Rev. D 90, 064038
\bibitem{sz19}C. Skordis and T. Zlosnik, 2019, Gravitational alternatives to dark matter with tensor mode speed equaling the speed of light. Phys. Rev. D 100, 104013
\bibitem{milgrom19a}M. Milgrom, 2019, Noncovariance at low accelerations as a route to MOND. Phys. Rev. D 100, 084039
\bibitem{dambrosio20}F. D'Ambrosio, M. Garg, and L. Heisenberg, 2020, Non-linear extension of non-metricity scalar for MOND.
    Phys. Lett. B  811, 135970
\bibitem{sz21}C. Skordis and T. Zlosnik, 2021, A new relativistic theory for Modified Newtonian Dynamics,
    Phys. Rev. Lett. 127, 161302
\bibitem{milgrom22}M. Milgrom, 2022, Broader view of bimetric MOND. Phys. Rev. D 106, 084010
\bibitem{milgrom10}M. Milgrom, 2010, Quasi-linear formulation of MOND. Mon. Not. R. Astron. Soc. 403, 886
\bibitem{milgrom23b}M. Milgrom, 2023, Tripotential MOND theories. Phys. Rev. D 108, 063009
\bibitem{milgrom23a}M. Milgrom, 2023, Generalizations of Quasilinear MOND (QUMOND).  Phys. Rev. D 108, 084005
\bibitem{tiret08}O. Tiret and F. Combes, 2008, Interacting Galaxies with Modified Newtonian Dynamics. Astron. Soc. Pacif. 396, 259
\bibitem{candlish15}G.N. Candlish, R. Smith, and M. Fellhauer, 2015, RAyMOND: an N-body and hydrodynamics code for MOND. Mon. Not. R. Astron. Soc. 446, 1060
\bibitem{thomas17}G.F. Thomas, B. Famaey, R. Ibata, F. Lueghausen, and P. Kroupa, 2017, Stellar streams as gravitational experiments. I. The case of Sagittarius. Astron. Astrophys. 603, A65
\bibitem{bilek17}M. B\'{\i}lek, I. Thies, P. Kroupa, and B. Famaey, 2018, MOND simulation suggests an origin for some peculiarities in the Local Group. Astron. Astrophys. 614, A59
\bibitem{banik22}I. Banik, I. Thies, R. Truelove, G. Candlish, B. Famaey, M.S. Pawlowski, R. Ibata, and P. Kroupa, 2022, 3D hydrodynamic simulations for the formation of the Local Group satellite planes. Mon. Not. R. Astron. Soc. 513, 129
\bibitem{banik22a}I. Banik, S.T. Nagesh, H. Haghi, P. Kroupa, and HS. Zhao, 2022, Overestimated inclinations of Milgromian disc galaxies: The case of the ultradiffuse galaxy AGC 114905. Mon. Not. R. Astron. Soc. 513, 3541
\bibitem{milgrom86}M. Milgrom, 1986, Solutions for the Modified Newtonian Dynamics Field Equation, Astrophys. J. 302, 617
\bibitem{brada00}R. Brada and M. Milgrom, 2000, The Modified Dynamics is Conducive to Galactic Warp Formation. Astrophys. J. Lett. 531, L21
\bibitem{brada00a}R. Brada and M. Milgrom, 2000, Dwarf Satellite Galaxies in the Modified Dynamics. Astrophys. J. 541, 556
\bibitem{milgrom15a}M. Milgrom, 2015, MOND theory. Canad. J. Phys. 93, 107
\bibitem{ignatiev07}A. Yu. Ignatiev, 2007, Is Violation of Newton’s Second Law Possible? Phys. Rev. Lett. 98, 101101
\bibitem{magueijo06}J. Bekenstein and J. Magueijo, 2006, Modified Newtonian dynamics habitats within the Solar System.
    Phys. Rev. D 73, 103513
\bibitem{magueijo12}J. Magueijo and A. Mozaffari, 2012, Case for testing modified Newtonian dynamics using LISA pathfinder.
    Phys. Rev. D 85, 043527
\bibitem{penner20}A.R. Penner, 2020, A proposed experiment to test gravitational anti-screening and MOND using Sun-Gas giant saddle points. Astrophys. Sp. Sc. 365, 154
\bibitem{wu15}X. Wu and P. Kroupa, 2015, Galactic rotation curves, the baryon-to-dark-halo-mass relation and space-time scale invariance. Mon. Not. R. Astron. Soc. 446, 330
\bibitem{hees16}A. Hees, B. Famaey, G.W. Angus, and G. Gentile, 2016, Combined Solar system and rotation curve constraints on MOND. Mon. Not. R. Astron. Soc. 455, 449
\bibitem{haghi16}H. Haghi, A.E. Bazkiaei, A.H. Zonoozi, and P. Kroupa, 2016, Declining rotation curves of galaxies as a test of gravitational theory. Mon. Not. R. Astron. Soc. 458, 4172
\bibitem{chae20}K.H. Chae et al., 2020, Testing the Strong Equivalence Principle: Detection of the External Field Effect in Rotationally Supported Galaxies. Astrophys. J. 904, 51
\bibitem{chae21}K.H. Chae et al., 2021, Testing the Strong Equivalence Principle. II. Relating the External Field Effect in Galaxy Rotation Curves to the Large-scale Structure of the Universe. Astrophys. J. 921, 104
\bibitem{stiskalek23}R. Stiskalek and H. Desmond, 2023, On the fundamentality of the radial acceleration relation for late-type galaxy dynamics, Mon. Not. R. Astron. Soc. 525, 6130
\bibitem{desmond23}H. Desmond, 2023, The underlying radial acceleration relation. Mon. Not. R. Astron. Soc. 526, 3342
\bibitem{hernandez12}X. Hernandez,  M.A. Jiménez, and C. Allen, 2012, Wide binaries as a critical test of classical gravity. Europ. Phys. J.  72, 1884
\bibitem{bps21}I. Banik, Indranil, C. Pittordis, and W. Sutherland, 2021, Detailed numerical implementation of the wide binary test. arXiv:2109.03827
\bibitem{chae23}K.H. Chae, 2023, Breakdown of the Newton-Einstein Standard Gravity at Low Acceleration in Internal Dynamics of Wide Binary Stars. Astrophys. J. 952, 128
\bibitem{hernandez23}X. Hernandez, 2023, Internal kinematics of GAIA DR3 wide binaries: anomalous behaviour in the low acceleration regime. Mon. Not. R. Astron. Soc. 525, 1401
\bibitem{chae23a}K.H. Chae, 2023, Robust Evidence for the Breakdown of Standard Gravity at Low Acceleration from Statistically Pure Binaries Free of Hidden Companions. arXiv:2309.10404
\bibitem{hernandez23a}X. Hernandez, V. Verteletskyi, L. Nasser, and A. Aguayo-Ortiz, 2023, Statistical analysis of the gravitational anomaly in {\it Gaia} wide binaries. arXiv:2309.10995
\bibitem{banik23}I. Banik, C. Pittordis, W. Sutherland, B. Famaey, R. Ibata, S. Mieske, and HS. Zhao, 2023, Strong constraints on the gravitational law from Gaia DR3 wide binaries. Mon. Not. R. Astron. Soc. in press
\bibitem{thomas18}G.F. Thomas et al., 2018, Stellar streams as gravitational experiments. II. Asymmetric tails of globular cluster streams.
   Astron. Astrophys. 609, A44
\bibitem{kroupa22}P. Kroupa et al., 2023, Asymmetrical tidal tails of open star clusters: stars crossing their cluster's pr\'{a}h challenge Newtonian gravitation. Mon. Not. R. Astron. Soc. 517, 3613
\bibitem{milgrom09a}M. Milgrom, 2009, MOND effects in the inner Solar system. Mon. Not. R. Astron. Soc. 399,  474
\bibitem{novak11}L. Blanchet and J. Novak, 2011, External field effect of modified Newtonian dynamics in the Solar system. Mon. Not. R. Astron. Soc. 412, 2530
\bibitem{bienayme14}O. Bienaym\'{e} et al., 2014,  Weighing the local dark matter with RAVE red clump stars. Astron. \& Astrophys. 571, A92.
\bibitem{loebman14}S.R. Loebman et al., 2014,  The Milky Way Tomography with Sloan Digital Sky Survey. V. Mapping the Dark Matter Halo.
  Astrophys. J. 794, 151.
\bibitem{angus15}G.W. Angus et al., 2015,  Mass models of disc galaxies from the DiskMass Survey in modified Newtonian dynamics.  Mon. Not. R. Astron. Soc. 451, 3551
\bibitem{milgrom15}M. Milgrom, 2015,  Critical take on "Mass models of disk galaxies from the DiskMass Survey in MOND".  arXiv:1511.08087.
\bibitem{angus16}G.W. Angus, 2016, The dynamics of face-on galaxies in MOND.
    J. Phys. 718, 032001
\bibitem{milgrom18}M. Milgrom, 2018, MOND and the dynamics of NGC 628, arXiv:1802.05140
\bibitem{lisanti19}M. Lisanti et al., 2019,  Testing dark matter and modifications to gravity using local Milky Way observables.
    Phys. Re. D 100, 083009.
\bibitem{milgrom94}M. Milgrom, 1994, Dynamics with a Nonstandard Inertia-Acceleration Relation: An Alternative to Dark Matter in Galactic Systems. Ann.  Phys. 229, 384
\bibitem{milgrom99}M. Milgrom, 1999, The modified dynamics as a vacuum effect. Phys. Lett. A 253, 273
\bibitem{milgrom02}M. Milgrom, 2002, MOND—theoretical aspects. New Astron. Rev. 46, 741
\bibitem{milgrom06}M. Milgrom, 2006, MOND as modified inertia. EAS Pub. Ser. 20, 217; arXiv:astro-ph/0510117
\bibitem{milgrom11}M. Milgrom, 2011,  MOND--particularly as modified inertia.  Act. Phys. Polon. B 42, 2175; arXiv:1111.1611
\bibitem{milgrom22a}M. Milgrom, 2022, Models of modified-inertia formulation of MOND. Phys. Rev. D 106, 064060
\bibitem{wang10}L.J. Wang, 2010, Recovering modified Newtonian dynamics by changing inertia. arXiv:1011.3618
\bibitem{namouni15}F. Namouni, 2015, Towards an interpretation of MOND as a modification of inertia. Mon. Not. R. Astron. Soc. 452, 210
\bibitem{alzain17}M. Alzain, 2017, Modified Newtonian Dynamics (MOND) as a Modification of Newtonian Inertia. J. Astrophys. Astron. 38, 59
\bibitem{cfp19} R. Costa, G. Franzmann, and J.P. Pereira, 2019, A local Lagrangian for MOND as modified inertia. arXiv:1904.07321
\bibitem{milgrom14b}M. Milgrom, 2014, General virial theorem for modified-gravity MOND. Phys. Rev. D 89, 024016
\bibitem{milgrom20a}M. Milgrom, 2020, The $a_0$ -- cosmology connection in MOND. arXiv:2001.09729
\bibitem{cushing81}J.T. Cushing, 1981, Electromagnetic mass, relativity, and the Kaufmann experiments. Am. J. Phys.e 49, 1133
\bibitem{milgrom06a}M. Milgrom, 2006, Massive particles in acoustic space-times: Emergent inertia and passive gravity. Phys. Rev. D 73, 084005
\bibitem{tiret07}O. Tiret,  F. Combes, G.W. Angus, G. W. , B. Famaey, and H.S. Zhao, 2007, Velocity dispersion around ellipticals in MOND. Astron. Astrophys. 476, L1
\bibitem{muller19}O. M\"{u}ller, B. Famaey, and HS. Zhao, 2019, Predicted MOND velocity dispersions for a catalog of ultra-diffuse galaxies in group environments. Astron. Astrophys. 623, A36,
\bibitem{haghi19}H. Haghi, P. Kroupa, I. Banik, X.n Wu, A. Hasani Zonoozi, B. Javanmardi, A. Ghari, O. Müller, J. Dabringhausen, and HS. Zhao, 2019, A new formulation of the external field effect in MOND and numerical simulations of ultra-diffuse dwarf galaxies -- application to NGC 1052-DF2 and NGC 1052-DF4. Mon. Not. R. Astron. Soc. 487, 2441
\bibitem{asencio22}E. Asencio et al., 2022, The distribution and morphologies of Fornax Cluster dwarf galaxies suggest they lack dark matter. Mon. Not. R. Astron. Soc. 515, 2981
\bibitem{bz15}I, Banik and H.S. Zhao, 2015, The External Field Dominated Solution In QUMOND \& AQUAL: Application To Tidal Streams.  	arXiv:1509.08457
\bibitem{chaemil22}K.H. Chae and M. Milgrom 2022, Numerical Solutions of the External Field Effect on the Radial Acceleration in Disk Galaxies. Astrophys. J. 928, 24
\bibitem{paucoklacka16}R. Pau\u{c}o and J.  Kla\u{c}ka, 2016, Sedna and the cloud of comets surrounding the Solar System in Milgromian dynamics. Astron. Astrophys. 589, A63
\bibitem{pauco17}R. Pau\u{c}o, 2017, Towards an explanation of orbits in the extreme trans-Neptunian region: The effect of Milgromian dynamics. Astron. Astrophys. 603, A11
\bibitem{migaszewski23}C. Migaszewski, 2023, On the origin of extreme trans-Neptunian objects within Modified Newtonian Dynamics. Mon. Not. Roy. Astron. Soc. 525, 805
\bibitem{mathur23}K. Brown and H. Mathur, 2023, Modified Newtonian Dynamics as an Alternative to the Planet Nine Hypothesis. Astron. J. 166, 168
\bibitem{ss06}F.J. Sánchez-Salcedo, J. Reyes-Iturbide, and X. Hernandez, 2006, An extensive study of dynamical friction in dwarf galaxies: the role of stars, dark matter, halo profiles and MOND.  Mon. Not. Roy. Astron. Soc. 370, 1829
\bibitem{tiret07a}O. Tiret and F. Combes, 2007, Evolution of spiral galaxies in modified gravity. Astron. Astrophys. 464, 517
\bibitem{nipoti08}C. Nipoti, L. Ciotti, J. Binney, and P. Londrillo, 2008, Dynamical friction in modified Newtonian dynamics. Mon. Not. Roy. Astron. Soc. 386, 2194
\bibitem{ss09}F.J. Sánchez-Salcedo, 2009, Gaseous drag on a gravitational perturber in Modified Newtonian Dynamics and the structure of the wake. Mon. Not. Roy. Astron. Soc. 392, 1573
\bibitem{angus09}G.W. Angus and A. Diaferio, 2009, Resolving the timing problem of the globular clusters orbiting the Fornax dwarf galaxy. Mon. Not. Roy. Astron. Soc. 396, 887.
\bibitem{vakili17}H. Vakili, P. Kroupa, and S. Rahvar, 2017, Type I Shell Galaxies as a Test of Gravity Models. Astrophys. J. 848,  55
\bibitem{bilek21}M. Bílek, HS. Zhao, B. Famaey, O. Müller, P. Kroupa, and R. Ibata, 2021,
 Evolution of globular-cluster systems of ultra-diffuse galaxies due to dynamical friction in MOND gravity. Astron. Astrophys 653, A170
\bibitem{bilek23}M. Bílek, HS. Zhao, B. Famaey, S.T. Nagesh, F. Combes, O. Müller, M. Hilker, P. Kroupa, and R. Ibata, 2023, Do old globular clusters in low mass galaxies disprove modified gravity? arXiv:2307.03202
\bibitem{vieira22}K. Vieira, et al. 2022, Milky Way Thin and Thick Disk Kinematics with Gaia EDR3 and RAVE DR5. Astrophys. J. 932, 28
\bibitem{sj06}M. Sereno,and Ph. Jetzer, 2006, Dark matter versus modifications of the gravitational inverse-square law: results from planetary motion in the Solar system. Mon. Not. R. Astron. Soc. 371, 626

\end{thebibliography}
\end{document}